\documentclass[fleqn,usenatbib]{mnras}

\usepackage[T1]{fontenc}

\DeclareRobustCommand{\VAN}[3]{#2}
\let\VANthebibliography\thebibliography
\def\thebibliography{\DeclareRobustCommand{\VAN}[3]{##3}\VANthebibliography}

\usepackage{graphicx}	
\usepackage{amsmath}	
\usepackage{amssymb}	

\usepackage{newtxtext,newtxmath}

\defcitealias{Bizyaev2014}{B14}
\defcitealias{Bizyaev2009}{BM9}
\defcitealias{Salo2015}{S15}
\defcitealias{Comeron2018}{C18}
\defcitealias{Mosenkov2010}{M10}
\defcitealias{Mosenkov2018}{M18}
\defcitealias{Bianchi2007}{B7}
\defcitealias{deGeyter2014}{DG14}
\defcitealias{Peters2017}{P17}
\defcitealias{Xilouris1999}{X99}
\defcitealias{Savchenko_2020}{S20}

\title[Deep imaging of spiral structure]{A multiwavelength study of spiral structure in galaxies. II. Spiral arms in deep optical observations}
\author[A. Mosenkov et al.]{
Aleksandr V. Mosenkov,$^{1}$\thanks{E-mail: aleksandr\_mosenkov@byu.edu}
Andrey D. Panasyuk,$^{2,3}$
Savanah Turner,$^{1}$
Crystal-Lynn Bartier,$^{1}$
\newauthor
Maria N. Skryabina,$^{3}$
Alexander A. Marchuk,$^{2,3}$
Sergey S. Savchenko,$^{2,3,4}$
Jakob Bergstedt,$^{5}$
\newauthor
Vladimir P. Reshetnikov,$^{2,3}$
and Ilia V. Chugunov$^{2,3}$
\\
\\
$^{1}$Department of Physics and Astronomy, N283 ESC, Brigham Young University, Provo, UT 84602, USA\\
$^{2}$Central (Pulkovo) Astronomical Observatory, Russian Academy of Sciences, Pulkovskoye Chaussee 65/1, St Petersburg 196140, Russia\\
$^{3}$Saint Petersburg State University, Universitetskij pr. 28, St. Petersburg 198504, Russia\\
$^{4}$Special Astrophysical Observatory, Russian Academy of Sciences, 369167, Nizhnii Arkhyz, Russia\\
$^{5}$Department of Physics, 525 S Center St, Brigham Young University-Idaho, Rexburg, ID 83460, USA
}

\date{Accepted XXX. Received YYY; in original form ZZZ}

\pubyear{2023}

\begin{document}
\label{firstpage}
\pagerange{\pageref{firstpage}--\pageref{lastpage}}
\maketitle

\begin{abstract}
In this paper, we look to analyse the spiral features of grand-design, multiarmed, and flocculent spiral galaxies using deep optical imaging from DESI Legacy Imaging Surveys. We explore the resulting distributions of various characteristics of spiral structure beyond the optical radius, such as the distributions of azimuthal angle, the extent of spiral arms, and of the spiral arm widths for the aforementioned galaxy classes. We also compare the measured properties for isolated galaxies and galaxies in groups and clusters. We find that, on average, compared to multiarmed and flocculent spiral galaxies, the spiral arms of grand-design galaxies exhibit slightly larger azimuthal angles, greater extent, and larger widths in the periphery of the galaxy. Furthermore, on average, isolated galaxies tend to have slightly smaller widths of outer spiral arms compared to galaxies in tight environments, which is likely related to the tidally-induced mechanism for generating wider outer spiral arms. We also report that breaks of the disc surface brightness profiles are often related to the truncation of spiral arms in galaxies. 
\end{abstract}




\section{Introduction}
\par
Spiral galaxies are the most abundant type of non-dwarf galaxies in the local Universe \citep{Conselice_2006}. This numerous type of disc galaxies demonstrates ongoing star formation, seen particularly well in their spiral arms, the appearance of which, along with the brightness of the bulge, serves for classifying spiral galaxies into different morphological subtypes \citep[see e.g.][and references therein]{1959HDP....53..275D,2005ARA&A..43..581S}. The connection of the spiral structure to the general galaxy characteristics and structural parameters of individual galaxy components is of great importance for our understanding of the formation mechanisms and evolution of spiral structures \citep{2015ApJ...802L..13D,2017MNRAS.471.2187D,2019ApJ...873...85D,2019ApJ...871..194Y,2022AstL...48..644R}. This, in its turn, is crucial to understanding the larger picture of general galactic evolution \citep{2014PASA...31...35D,Sellwood_2022}.
\par
Spirals in galaxies can be morphologically classified into three main classes: grand design, multiarmed, and flocculent \citep{1990NYASA.596...40E}. Grand-design galaxies are known for their two highly symmetric and continuous spiral arms, whereas the spiral arms in flocculent galaxies are closely spaced and fragmented, resulting in a fleece-like appearance. Multiarmed galaxies, as the name suggests, are spiral galaxies with more than two defined arms, which often show bifurcations with a feather-like structure \citep{1987ApJ...314....3E}. Within these three classes, spiral galaxies exhibit differing morphological characteristics \citep{Conselice_2006}.
\par
The literature sets forth major differing mechanisms as to how spiral structure in galaxies forms. From the obvious visible connection, many studies have suggested that spiral arms are driven by bars \citep{Toomre_1969,Feldman_1973,Kormendy_1979}, although it was shown that the spiral arms of some barred galaxies do not have the same pattern speed \citep{Sellwood_1988,2009AJ....137.4487B,Lieb_2021}. Many numerical simulations demonstrate that spiral arms may be driven by a direct response of the bar forcing  \citep{1981ApJ...247...77S,1985A&A...150..327C,1994PASJ...46..165W,2012MNRAS.426..167G,2009ApJ...706..471B}. Developed for its explanation, the manifold theory also finds significant support in simulations \citep{2010MNRAS.407.1433A,2012MNRAS.426L..46A}. However, the examination of observational evidence yields contradictory results on the triggering of spiral arms by bars \citep{1998MNRAS.299..685S,2004AJ....128..183B,2010ApJ...715L..56S,2011MNRAS.414..538K}.
\par
A second possible explanation for the formation of spiral arms in galaxies is self-excitation, either by quasi-steady, global modes in the stellar disc (also known as density waves, \citealp{1964ApJ...140..646L,1989ApJ...338..104B}), or by local, transient self-gravitational instabilities (also known as dynamic spirals, \citealp{1964ApJ...139.1217T,2011MNRAS.410.1637S}). Both theories are well developed and have found a convincing set of supporting evidence (see the review in \citealp{2014PASA...31...35D,Sellwood_2022,2016ARA&A..54..667S}). However, dynamic spirals are much easier to obtain in numerical simulations, whereas density waves are difficult to reproduce \citep{2003MNRAS.344..358B,2011ApJ...730..109F,2013ApJ...763...46B,2013ApJ...766...34D}. In the so-called swing amplification mechanism \citep{1965MNRAS.130..125G,1966ApJ...146..810J,1978ApJ...222..850G,1981seng.proc..111T,1991dodg.conf..341T}, a density enhancement, formed by self-gravity, is stretched out by differential rotation of the disc. This mechanism plays an important role in the foundations of both aforementioned theories \citep{2011ApJ...730..109F,1984ApJ...282...61S,1985ApJ...298..486C,2011MNRAS.410.1637S,2003MNRAS.344..358B,2013ApJ...766...34D,2000Ap&SS.272...31S}.
\par
Finally, another potential inducing mechanism for spiral arms is tidal interactions with neighbouring galaxies. In fact, some have argued that all spiral structures are driven by such a mechanism \citep{1979ApJ...233..539K,2003MNRAS.344..358B}. Moreover, well-seen grand-design spirals form in many numerical studies of interacting galaxies \citep[see e.g.][]{2000MNRAS.319..377S,2010MNRAS.409..396D,2008ApJ...683...94O,2011MNRAS.414.2498S}. In these works, a perturber galaxy has the greatest impact when orbiting in the plane of the main galaxy \citep{1972ApJ...178..623T}. Furthermore, simulations have shown that tidal forces can also trigger the formation of a bar, linking these two formation mechanisms \citep{Peschken_2019}. It is worth noting that tidal excitation could also be induced not only by visible companions, but also by orbiting clumps of dark matter \citep{2006ARep...50..785T,2009ApJ...697..293D}. 

Before a consensus can be reached on which mechanisms contribute the most to the formation of spiral structure, more data must be obtained to analyse the characteristics of the observed spiral arms in galaxies, which has been proven to not be an easy task \citep{2010MNRAS.409..396D}.
\par
To mitigate the difficulty of this task, surveys of spiral structure characteristics could help narrow down the proposed mechanisms of spiral arm formation in galaxies \citep[see e.g.][]{1981AJ.....86.1847K,1998MNRAS.299..672S,2011ApJ...737...32E,2011MNRAS.414..538K,2018ApJ...869...29Y,2019A&A...631A..94D,Savchenko_2020,2020ApJ...900..150Y}. However, describing the morphology of galaxies is known to be a difficult and often subjective task. 
The appearance of spiral arms can be strongly affected by both the wavelength in which they are observed \citep{2002ApJS..143...73E,2015ApJS..217...32B} and the photometric depth of the observation \citep[see e.g.][and references therein]{2023A&A...671A.141M}. For example, a recent study by \citet{2023MNRAS.525.3016M} revealed that the ring galaxy UGC\,4599 hosts an extremely faint outer disc with an embedded spiral structure out to $\approx45$~kpc, whereas the blue ring, which determines the edge of the galaxy in ordinary SDSS exposures with a surface brightness depth of 26.5~mag\,arcsec$^{-2}$, has a modest radius of $\approx9$~kpc.
Attempts have been made to automate galactic classification in order to process large amounts of data faster and more consistently, including asymmetric features such as spiral arms (see, e.g., \citealt{2023MNRAS.518.1022S}). 
However, these methods have not become advanced enough to describe the complex structure of spiral arms in galaxies and retrieve their parameters. Thus, it remains necessary to develop consistent and well-defined methods of characterising the galactic spiral structure available in the literature, as well as to propose new approaches to quantify the properties of spiral arms in great detail. The latter is especially important for the galaxy outskirts, where standard methods (such as, for example, a Fourier analysis \citealt{1988A&AS...76..365C,2020ApJ...900..150Y}) usually fail to trace deviating outer arms in spiral galaxies.
\par
Recently, \citet{2019MNRAS.490.1539R} obtained deep observations for a sample of 119 nearby galaxies. Using their results, we estimate the ratio $R_{28}/R_{25}=2.31\pm1.19$ for spiral galaxies, where $R_{28}$ is the radius of an isophote at 28~mag$\,$arcsec$^{-2}$ level in the $r$ band and $R_{25}$ is the optical radius of the isophote at 25~mag$\,$arcsec$^{-2}$, as taken from HyperLeda\footnote{\url{http://leda.univ-lyon1.fr/}} \citep{Makarov2014}. As we can see from this ratio, a significant region of the galaxy can be traced beyond the optical radius where spiral arms may, in principle, manifest themselves, especially those induced by tidal interactions. 
\par
In this paper, we illustrate how deep images and carefully defined methods for quantitatively describing spiral arm features can be used to characterise a large number of galaxies. The first paper in this series (\citealt{Savchenko_2020}, hereafter \citetalias{Savchenko_2020}), focused on quantifying the general characteristics of spiral galaxies using optical observations from the Sloan Digital Sky Survey (SDSS, \citealt{York_2000}). The parameters of spiral structure used were the width, pitch angle, and the portion of the total galactic luminosity that came from the arms. In this study, we measure the same parameters (with some additional ones) for the spiral structure but beyond the optical radius $R_{25}$ (in \citetalias{Savchenko_2020} this was done for $\sim2$~mag~arcsec$^{-2}$ less deep SDSS photometry than what we utilise in this study). 
\par
In this paper, we predominantly focus on the radial dependence of pitch angle and arm width in the galaxy outskirts. Such a radial dependence has been poorly studied in the past due to the difficulty of observing a faint spiral structure far from the galactic centre. We use deep optical images from the Dark Energy Spectroscopic Instrument (DESI) Legacy Surveys (DESI, hereafter, \citealt{Dey_2019}) in conjunction with SDSS images. We show how deeper optical observations of galaxies allow for more complete radial spiral structure analysis than observations obtained from SDSS alone. 
\par
Another interesting question is the influence of spiral structure in a galaxy on the retrieved parameters of its main components (a disc and a bulge) when the presence of spiral arms is neglected. Moreover, the stellar discs in galaxies may have different types of the surface brightness profile \citep[see e.g.][and references therein]{2011ARA&A..49..301V}. It is important to investigate whether these different types of disc profile are related to the properties of spiral structure: the class of the arms, to dramatic changes in the pitch angle or the width of the arms, and to the extent of spiral arms. 
\par
The aims of our article include i) a photometric characterization of {\it outer} spiral structure in galaxies primarily using deep optical imaging and supplementary ultraviolet observations, ii) an analysis of the factors which may affect the morphology of spiral arms (such as, e.g. the spatial environment of the galaxy, secular evolution, etc.), iii) the relation of the observed breaks of the galaxy surface brightness profiles with the extent of spiral structure, and iv) the influence of the spiral structure on the retrieved parameters of the galaxy disc. Deep DESI photometry allows us to reach the set goals with relative ease. The sample of 155 galaxies from \citetalias{Savchenko_2020} is suitable for our needs and comprises spiral galaxies of various morphological types and classes of spiral structure. This study enables us to shed new light on the fundamental properties of the spiral structure in galaxies.
\par
The structure of this paper is as follows. In Section~\ref{sec:samples}, we briefly introduce our sample of galaxies and then detail the preparation of data for the galaxies in Section~\ref{sec:data}. Section~\ref{sec:methods} is concerned with the methods used to characterise the spiral arms of our sample galaxies and to obtain and fit their surface brightness profiles. We present our results in Section~\ref{sec:results}. The summary and conclusions are found in Section~\ref{sec:conclusion}. All magnitudes and surface brightnesses are given in the standard AB magnitude system.

\section{The sample}
\label{sec:samples}
The original sample used in this study is described in detail in \citetalias{Savchenko_2020}. Readers seeking a full description of the selection criteria used to generate the sample should refer to their article. In summary, the sample is composed of 155 relatively nearby spiral galaxies imaged with SDSS. The objects were originally chosen from the GalaxyZoo project sample \citep{Lintott_2008} and EFIGI \citep{Baillard_2011} catalogues. Only galaxies with optical diameters greater than 50~arcsec and at least one traceable spiral arm were retained in the sample: the angular resolution of these galaxies in the SDSS is sufficient to resolve the width of their spiral arms. The sample is dominated by grand-design and multiarmed galaxies with a little fraction of flocculent galaxies (the main reason for this is the difficulty in tracing individual spiral arms in this class of spirals). However, we can regard the selected galaxies as a quite representative sample for studying the general structure of spiral arms in galaxies. The sample comprises galaxies with absolute magnitudes $-19.5\lesssim M(r)\lesssim -23$ and is dominated by Hubble stages $2\lesssim T \lesssim 6$, although a handful of early-type and very late-type spirals are also present in the sample (see figure~1 in \citetalias{Savchenko_2020}).

All the objects in our sample can be considered face-on galaxies or galaxies with a moderate inclination (the average inclination for the sample is $\langle i \rangle =36\degr\pm13\degr$), so their spiral structure can be investigated in great detail. However, for uniformity, all galaxy images should be deprojected to a face-on view given the known position and inclination angles, which have been accurately estimated in \citetalias{Savchenko_2020} for SDSS images using outer galaxy isophotes with the spirals masked.

\section{The data}
\label{sec:data}

In our analysis of spiral structure, we use deep optical observations from DESI and the legacy of the Galaxy Evolution Explorer (GALEX, \citealt{Martin_2005}) satellite to compare the extent of the spirals and the properties of the stellar disc in these two adjacent spectral regions. Since the spiral arms in galaxies demonstrate enhanced star formation, ultraviolet and bluish optical bands usually trace them very well (although red spirals also exist; see e.g. \citealt{2010MNRAS.405..783M} --- we discuss this in Sect.~\ref{sec:results}). Thus, in spite of the apparent differences of the DESI and GALEX observations (as we discuss below), both sources should, in general, give similar results in the extent of (blue) spiral arms (which we demonstrate in Sect.~\ref{sec:outer_structure}).

The DESI Legacy Survey aims to create an inference model of the extragalactic sky in the optical and near-infrared using several relatively large telescopes with wide-field cameras. Specifically, DESI has imaged over 14,000 $\mathrm{deg}^2$ of the extragalactic sky with an average surface brightness depth of $28.5-29.5$~mag\,arcsec$^{-2}$ (as measured in a box of $10\arcsec \times 10\arcsec$ at a $3\sigma$~level -- we use the same definition of photometric depth throughout the text) in the $g$ band. Here, we use deep imaging obtained for each of the 155 galaxies as part of DESI and compare these observations to original SDSS photometry to explore whether the increased photometric depth can help us see faint spirals that are not seen or barely seen in SDSS images. For this, we use DESI photometry in the filters $g$ (photometric depth $29.02\pm0.22$~ mag\,arcsec$^{-2}$) and $r$ (photometric depth $28.38\pm0.25$~mag\,arcsec$^{-2}$) filters. To better trace the spiral arms in galaxies, we create an average image of two in the $g$ and $r$ wavebands. This is also justified by the fact that some spirals are bluer and others are redder, so that by using an averaged image we can trace spiral arms of different colours more consistently for different types of spirals.

Images of the sample galaxies obtained with GALEX are used to compare the appearance of spiral arms in ultraviolet and optical bands and the ``redness'' of spiral arms. 
Since GALEX FUV images have a lower photometric depth than those obtained in the NUV band, we decided to use the latter for mapping the relatively young stellar population tightly related to the star formation, which, in its turn, is linked to the spiral arms. Of our sample, 140 galaxies have GALEX NUV observations with a photometric depth of $29.10\pm1.19$~mag~arcsec$^{-2}$.

All data were retrieved from the corresponding databases and processed in a similar way. The sky background around the target galaxy was re-estimated within an annulus with a radius of three optical radii and a width of 1/3 optical radius with a polynomial of 2nd degree. For each galaxy image, an automatically created mask to exclude close or overlapping foreground and background sources was revisited and corrected if needed. Finally, all of the galaxy images were deprojected to face-on orientation (see Sect.~\ref{sec:samples}). Similar procedures were performed for SDSS images of the same galaxies in \citetalias{Savchenko_2020}.

\section{Methods}
\label{sec:methods}

\subsection{Fitting of spiral structure}
\label{sec:sp_structure}
\begin{figure*}
    \centering
    \includegraphics[width=0.99\linewidth]{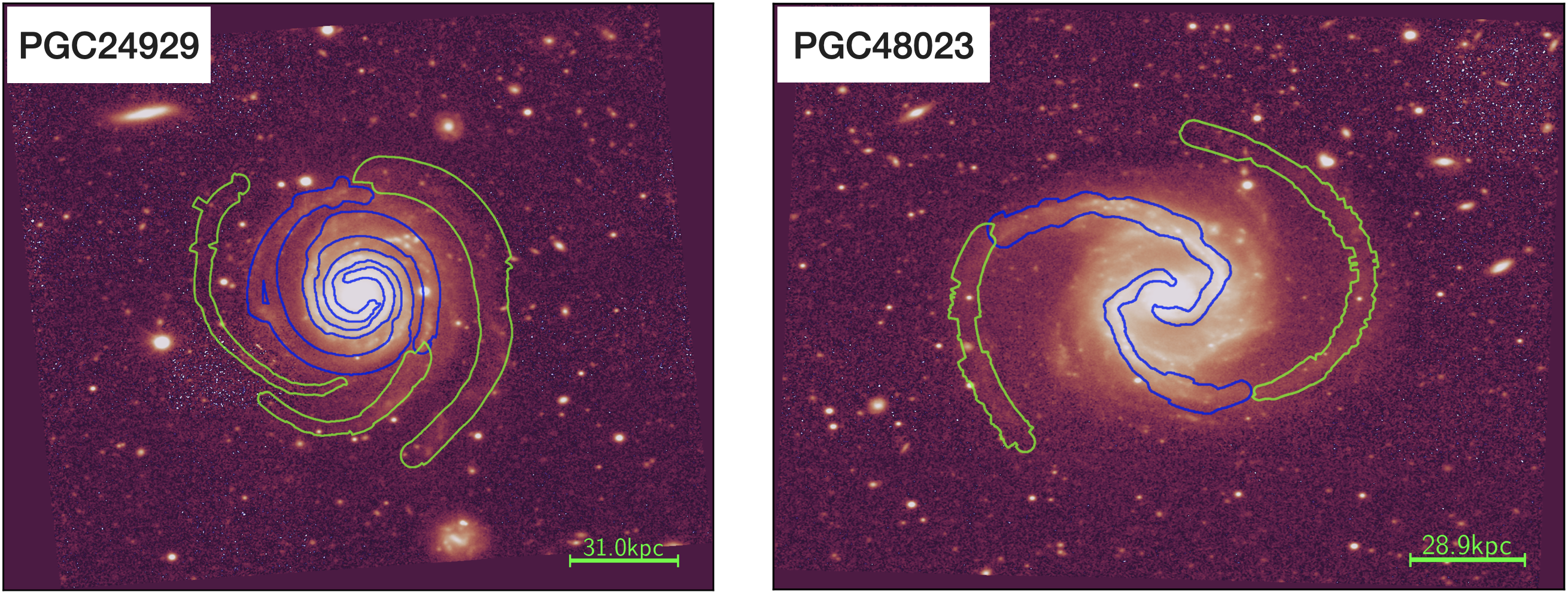}
    \caption{Examples of averaged $gr$ DESI images for two galaxies in our sample. Both images were deprojected to face-on orientation. The blue contours show masks of spiral arms traced by \citetalias{Savchenko_2020} using SDSS photometry, while the spiral structure outlined in green is traced using the deeper DESI photometry.}
    \label{fig:spiralmask}
\end{figure*}

The fitting of spiral structure in the sample galaxies was performed using averaged $gr$ DESI images. As noted above, this is done to i) increase the signal-to-noise ratio, which is of high importance in the low surface brightness outer region of a spiral galaxy, and ii) equally recover the shape of the blue and reddish arms in the periphery of the galaxy. In fact, \citetalias{Savchenko_2020} did not find any significant difference between the characteristics of the spiral structure in the $gri$ SDSS bands for galaxies in our sample, except for the arm fraction in total galaxy luminosity and the colour of the spiral structure, so this approach should not introduce any bias in our results.

A procedure for methodically characterising spiral arms is detailed in \citetalias{Savchenko_2020}. We employ the same method but use deeper images from DESI rather than SDSS observations. With the deeper images, spiral arm features are more easily traced beyond the optical radius, allowing for more complete analysis. Below we summarise details of the algorithm used, but refer the reader to \citetalias{Savchenko_2020} and \citet{2020RAA....20..120M} for a more complete procedural explanation.

The core idea of the method is to analyse the light distribution in photometric cuts made across spiral arms at different points of the spiral structure. To do so, we need a set of regions along each traceable spiral arm, so in the first step, we utilised the SAO DS9 package \citep{Joye2003}, to manually place circular regions covering each spiral arm to continue the mask of the spiral arms created in \citetalias{Savchenko_2020} for SDSS images. The radii of the regions were chosen to cover the whole width of the arm perpendicular to its direction at a given point, from the nearest inward to nearest outward interarm space. The arm was traced visually using the histogram equalization scale function in DS9 out to a radius where the spiral arm became indistinguishable from the background (see below how we selected the radius of the truncation for the traced spiral arms). To reduce the scattering of manually selected points, we then applied a moving window smoothing algorithm.

In the next step, the points on each arm were used to compute a set of perpendiculars to the arm at regular intervals and to make photometric cuts along these perpendiculars. Several subsequent cuts were averaged, and the resulted photometric profile was then fitted with an asymmetric Gaussian function given in eq.~\ref{eq:asymmetricGaussian}, where $w_1$ and $w_2$ are fit parameters describing the inner and outer arm half-widths, respectively, $I_0$ is the amplitude of the Gaussian profile and $r_\mathrm{peak}$ is the location of the brightness peak within each photometric cut: 

\begin{equation}
    I_\mathrm{model} = I_0 \times \begin{cases}
        \mathrm{exp}\left(-\frac{[r-r_\mathrm{peak}]^2}{w_1^2}\right), & r < r_\mathrm{peak} \\
        \mathrm{exp}\left(-\frac{[r-r_\mathrm{peak}]^2}{w_2^2}\right), & r > r_\mathrm{peak} \\
    \end{cases}
    \label{eq:asymmetricGaussian}
\end{equation}

We note that after this step, we refined the positions of the brightest points (calculated from the fitted values of $r_\mathrm{peak}$) along the arm forming an arm ridge line. Further analysis of the spiral arms utilises these new arm locations, not the manually placed regions. Slices were fitted (and stored in our results) until the peak intensity had become systematically lower than 3\,$\sigma$ of the standard deviation of the residual (i.e. the data minus the model).

To calculate the pitch angle of a spiral arm approximated with a logarithmic spiral, the distribution of the points on the arm ridge line was then fit with a linear regression model. The slope of the regression line was recorded as the tangent of the pitch angle for that arm. For each galaxy, we also computed an average pitch angle (as the simple mean for all spiral arms in the galaxy) and its variation with radius, 
 by fitting a linear regression model to individual pieces of the spiral structure as described in \citetalias{Savchenko_2020}. The same was done for the entire width of the spiral arms $w=w_1+w_2$. In addition, we estimated how the width of a spiral arm changes with radius, approximating this behaviour with a linear function with a slope $s$. We estimated all these parameters within and outside the optical radius $R_{25}$, and at all radii both in SDSS and DESI where we traced the spiral arms. 

The deep optical photometry from DESI indeed allows the spiral structure of the sample galaxies to be traced well beyond the optical radius, as illustrated in Fig.~\ref{fig:spiralmask} for two sample galaxies, namely PGC\,24929 and PGC\,48023. 

We also used the supplementary NUV observations to probe the extent of spiral structure based on young populations of stars that typically reside in spiral arms. Since the angular and image resolutions of the NUV images are quite poor for studying spiral structure in most of our sample galaxies, we decided to only determine the outer (maximum) radius of the well-defined spiral arms at a level of $3\,\sigma$ of the background. 

The results of our fit for the GALEX and DESI data can be found in the supplementary material in the electronic version of the MNRAS; the columns of the data table are described in Table~\ref{tab:spirals} in Appendix~\ref{sec:table_1}.

\subsection{Fitting of galaxy profiles}
\label{sec:profiles}

To investigate the one-dimensional surface brightness distribution in our galaxies, we obtained azimuthaly averaged profiles for each deprojected galaxy image in the NUV and $r$ bands (the mask of foreground and background objects was taken into account). These profiles demonstrate the presence of galaxy structural components dominating in the corresponding filter: a relatively young stellar population in the NUV band and a mixed stellar population in the $r$ band. In Sect.~\ref{sec:results}, we compare how these profiles in the two different spectral regions compare with each other and are related to the characteristics of spiral structure.

To determine the type of disc profile (see below), we use a piecewise linear function with zero to three breaks on the surface brightness (expressed in mag~arcsec$^{-2}$) distribution in the disc. We mask off the inner-galaxy region where a bulge, a bar, and/or an active galactic nucleus may reside. The presence of red/reddish central components is usually not apparent in the NUV because they are mainly dominated by Population II stars, but it is not negligible in the optical. Here, potential caveats of this approach include i) an underestimation of the scattered light, the outer profile of which may be misinterpreted as an up-bending disc profile in the periphery of the galaxy, and ii) the light from the central component(s) may influence the model of the disc. Although adding a central component in our fitting would result in a more reliable overall galaxy model, we decided to minimise the number of free parameters and focus on the disc parameters only. Also, the percentage of galaxies with a relatively high bulge-to-disc luminosity ratio should be low, since galaxies with Hubble stage earlier than Sbc constitute only $\approx33$\% of the sample (see figure~1 in \citetalias{Savchenko_2020}). Full-fledged decomposition of the galaxy images will be carried out in our future paper. 

The fitting of the galaxy profiles was conducted manually by fixing the number of breaks on the disc profile and specifying their first-guess locations. Although DESI galaxy profiles typically extend down to $\sim30$~mag~arcsec$^{-2}$ in the $r$ band, we do not consider surface brightness distributions beyond a level of $\sim 28$~mag~arcsec$^{-2}$ because i) the possible scattered light from the central source may dominate the profile at lower surface brightness levels (the wings of the PSF may appear right beyond this level, \citealt{2015A&A...577A.106S,2016ApJ...823..123T,2019MNRAS.490.1539R}) and ii) this part of the profile may be dominated by a stellar halo \citep[see e.g.][]{2016ApJ...830...62M,2020MNRAS.495.4570M,2022ApJ...932...44G}, iii) galaxies with a large angular diameter may suffer from sky under/over subtraction. As we will see below, almost all galaxies in our sample possess spiral arms that truncate within $\sim28$~mag~arcsec$^{-2}$, so our classification and fitting of the disc profile should not be affected by this truncation of the galaxy profile.

During our fitting, the model is optimised using a Nelder-Mead simplex algorithm to find the minimum of the $\chi^2$ function. We minimised the number of breaks based on the BIC values after adding a new break, but the maximum number of breaks was three --- this appeared to be sufficient to accurately fit galaxy profiles of a complex shape. The results of our fitting are provided in the supplementary material in the electronic version of the MNRAS; the columns of the data table are described in Table~\ref{tab:disks} in Appendix~\ref{sec:table_2}.

Our classification of the disc surface brightness profiles is performed following the literature. Since the seminal works by \citet{2005ApJ...626L..81E} and \citet{2006A&A...454..759P}, disc galaxies have been separated into three major classes (following and expanding upon the system devised by \citealt{Freeman1970}). Type I discs demonstrate surface brightness profiles described well by a single exponential decline. Type II discs are best described by a broken exponential with a steeper decline in the disc outskirts. Finally, Type III discs are characterised by a broken exponential with a shallower decline in the outskirts. Although Type II disc breaks are the most common \citep{2006A&A...454..759P,2008AJ....135...20E,2014MNRAS.441.1992L}, different modifications of Type II and Type III profiles are not rare \citep{2011AJ....142..145G,2018A&A...610A...5C}. Using the result of our fitting, all galaxy profiles were independently annotated in the NUV and $r$ bands. 

\begin{figure*}
    \centering
    \includegraphics[width=1.0\linewidth]{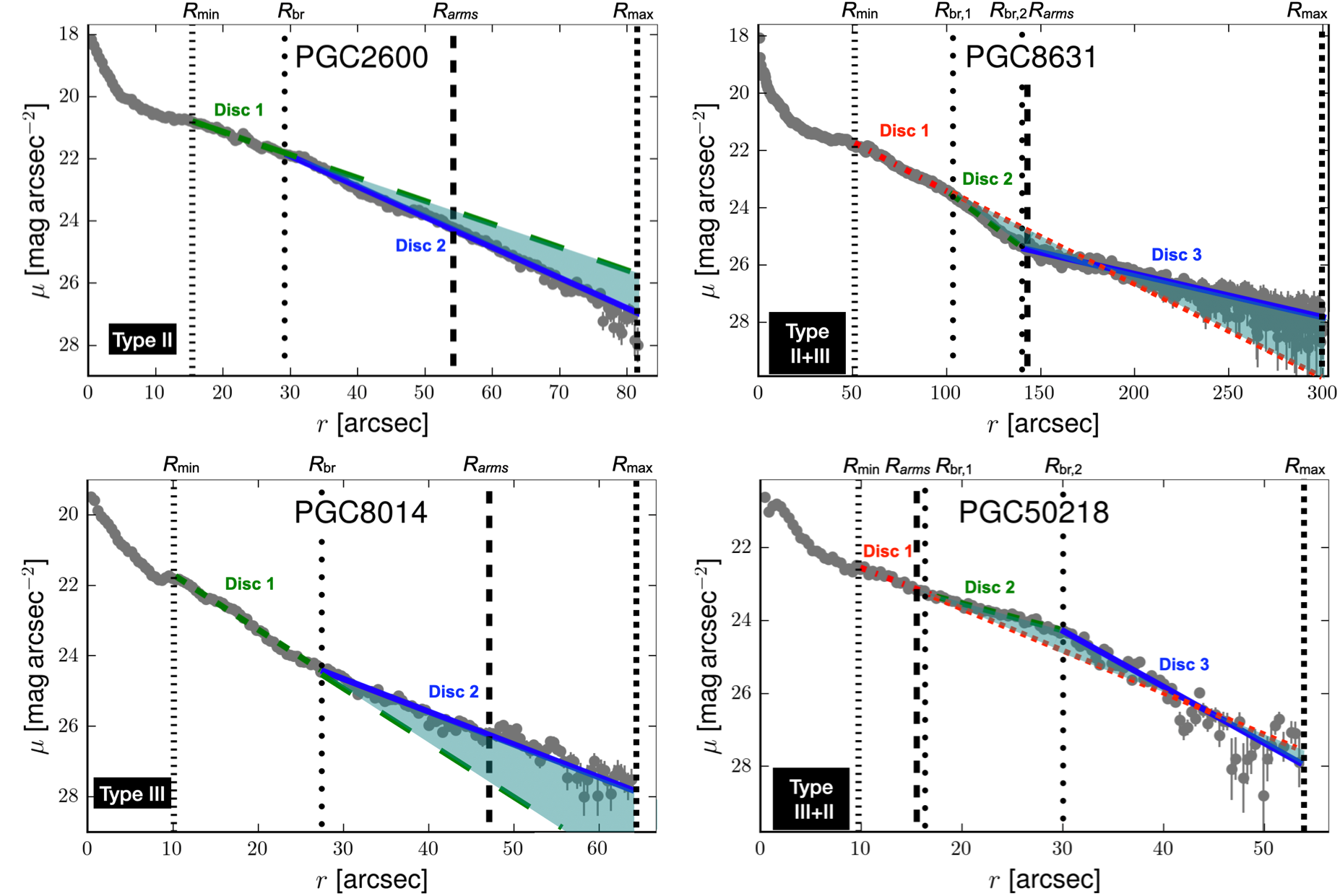}
    \caption{Sketches demonstrating results of our fitting for galaxies with different disc types. The shaded area is $|S_1-S|$ in eq.~\ref{eq:BS}. $R_\mathrm{arms}$ depicts the extent of the spiral arms in this galaxy. }
    \label{fig:SBprofiles_example}
\end{figure*}

In addition to the classification of the disc profiles, we quantify the bending strength of the disc profile using the following approach. We use the first (inner) model disc profile and compare the alignment of all model data points with respect to this first disc (it is extrapolated from the beginning of the disc profile, that is, the inner radius $R_\mathrm{min}$ where we start our disc fitting, down to the level $28$~mag~arcsec$^{-2}$ at $R_\mathrm{max}$). Then the bending strength ($BS$) of the galaxy profile, where we consider the disc, is computed as

\begin{equation}
    \mathrm{BS} = 100\,\frac{(S_1 - S)}{S}\,,
    \label{eq:BS}
\end{equation}

where $S_1$ is the area under the first model disc profile computed within the range [$R_\mathrm{min}$,$R_\mathrm{max}$] and $S$ is the area under the galaxy model profile within the same range of radii (see Fig.~\ref{fig:SBprofiles_example}) -- both areas are calculated for the profiles normalised by the central surface brightness of the first disc. From eq.~\ref{eq:BS}, it is obvious that galaxies with Type I profiles should have $BS=0$, whereas Type II profiles have strongly negative $BS$ values and Type III profiles demonstrate positive values of $BS$ (note that the above areas are found for the mag~arcsec$^{-2}$ profiles). The meaning of the $BS$ parameter is straightforward: it is the percentage of the {\it total} disc area covered by the extrapolated inner disc within the range [$R_\mathrm{min}$,$R_\mathrm{max}$] in the $r$--$\mu$ space. Note that the $BS$ indicator is only sensitive to the overall shape of the profile, so for profiles with multiple breaks, $BS$ accounts for the {\it overall} difference between the extrapolated inner disc profile and the outer disc profiles. This implies that the sign and value of the bending strength for multiple-break profiles depend on the overall up/down-bending degree for the entire disc profile. For example, if a Type II+III profile is dominated by the down-bending of the middle disc segment, the $BS$ value will be negative. However, if an outer up-bending segment makes a larger contribution to the bending strength than the middle down-bending profile, the $BS$ indicator will be positive. 

To be able to compare the $BS$ indicator in the NUV band with that in the $r$ band, we use the same range of radii when computing the BS indicator, as estimated in the $r$ band. We ensured that the DESI and NUV profiles indeed traced the beginning and end of the disc dominance in approximately the same range of radii on the galaxy profiles.

\section{Results and discussion}
\label{sec:results}
In this section, we present some general trends between the general characteristics of spiral structure and other galaxy properties for our sample galaxies with the help of the deep DESI images and complementary NUV observations. First, we make a comparison of the disc profiles in the optical and NUV, and we show how the extents of the spiral arms agree in these wavebands. Second, we consider a possible relation between the galaxy disc profile and the class of spiral structure and how these results depend on wavelength. We explore the outer spiral structure based on the results of our arm fitting in dependence of class of spiral arms and spatial environment of galaxies. Finally, we address the question of the importance of including spiral arms in the decomposition of galaxy images and how the retrieved parameters of the disc may be affected by the presence of a spiral structure. Throughout this section, we perform a linear regression analysis using a linear regression equation in the form $Y=aX+b$ with the Pearson coefficient $\rho$.

\subsection{Comparison of the results in the optical and NUV}
\label{sec:optical_vs_uv}
\begin{figure*}
    \centering
    \includegraphics[width=0.4\linewidth]{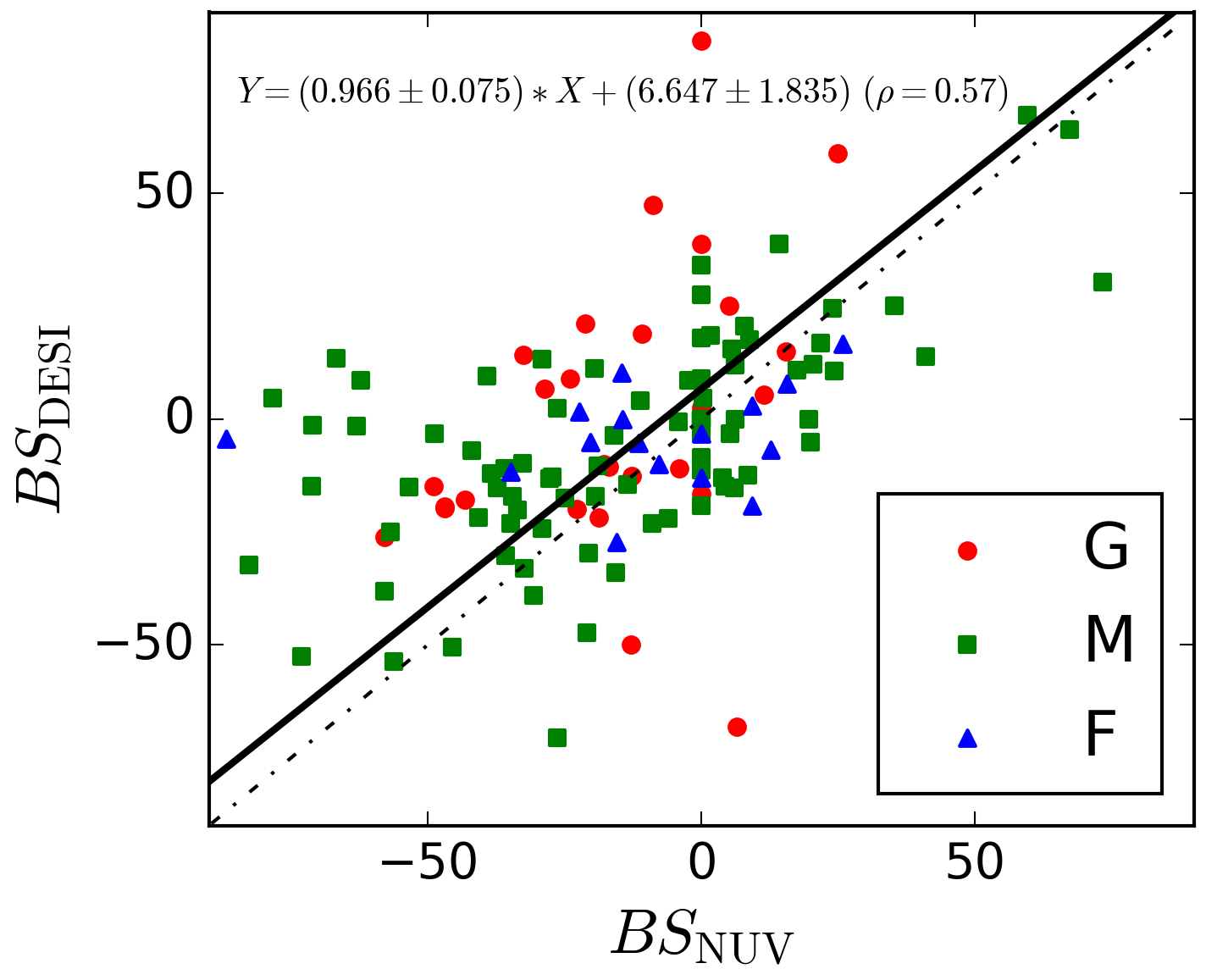}
    \includegraphics[width=0.4\linewidth, height=5.78cm]{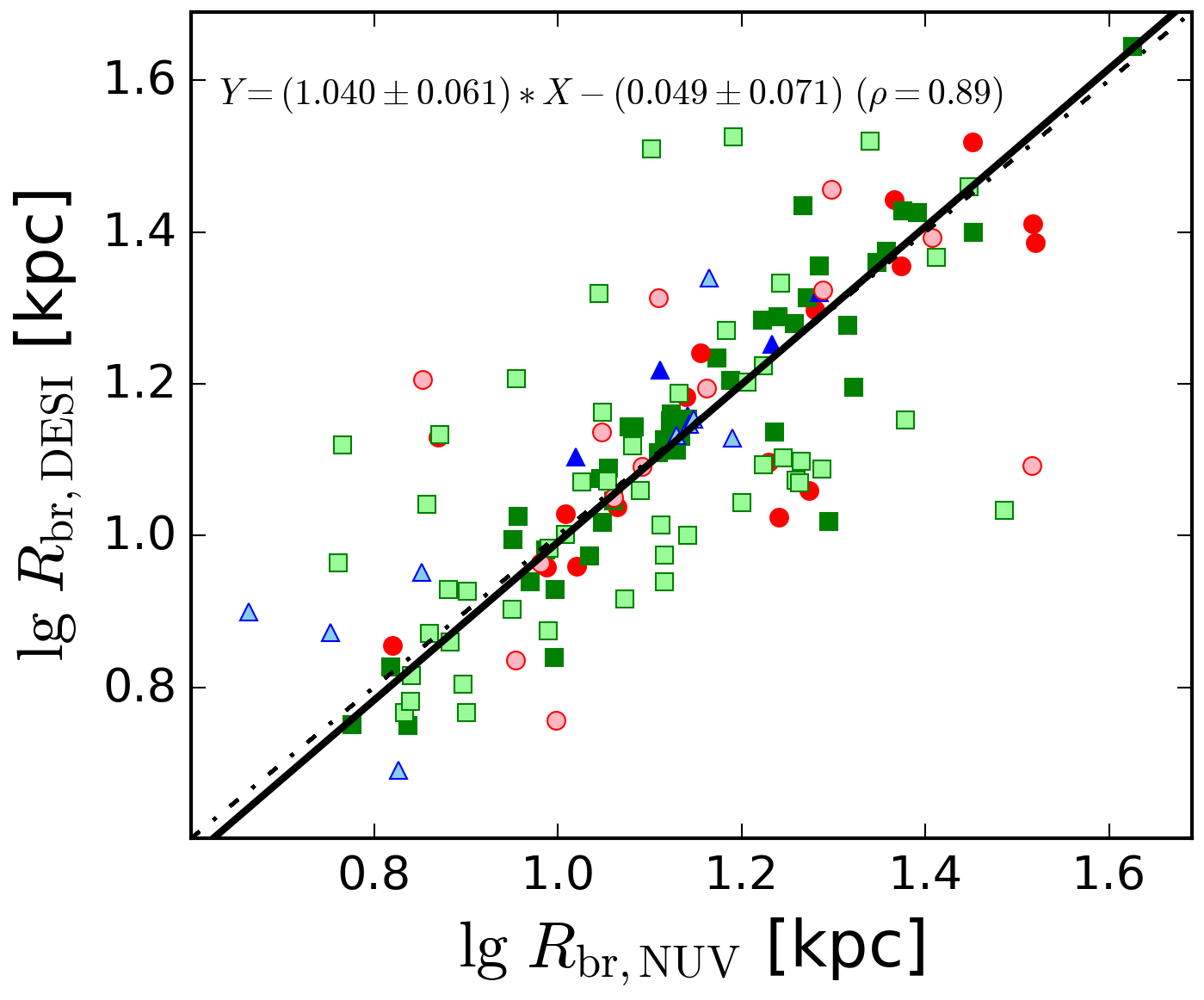}
    \includegraphics[width=0.4\linewidth]{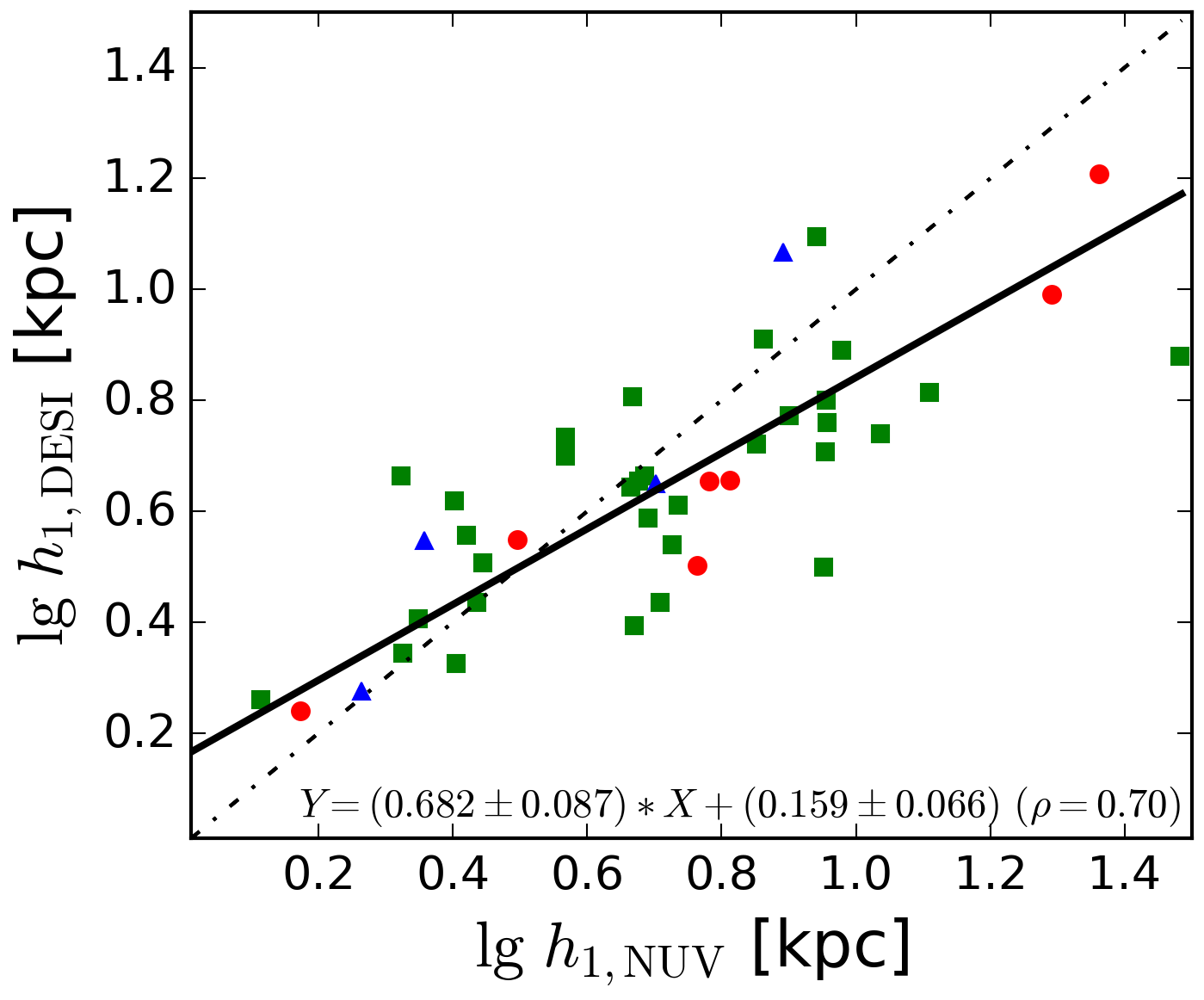}
    \includegraphics[width=0.4\linewidth]{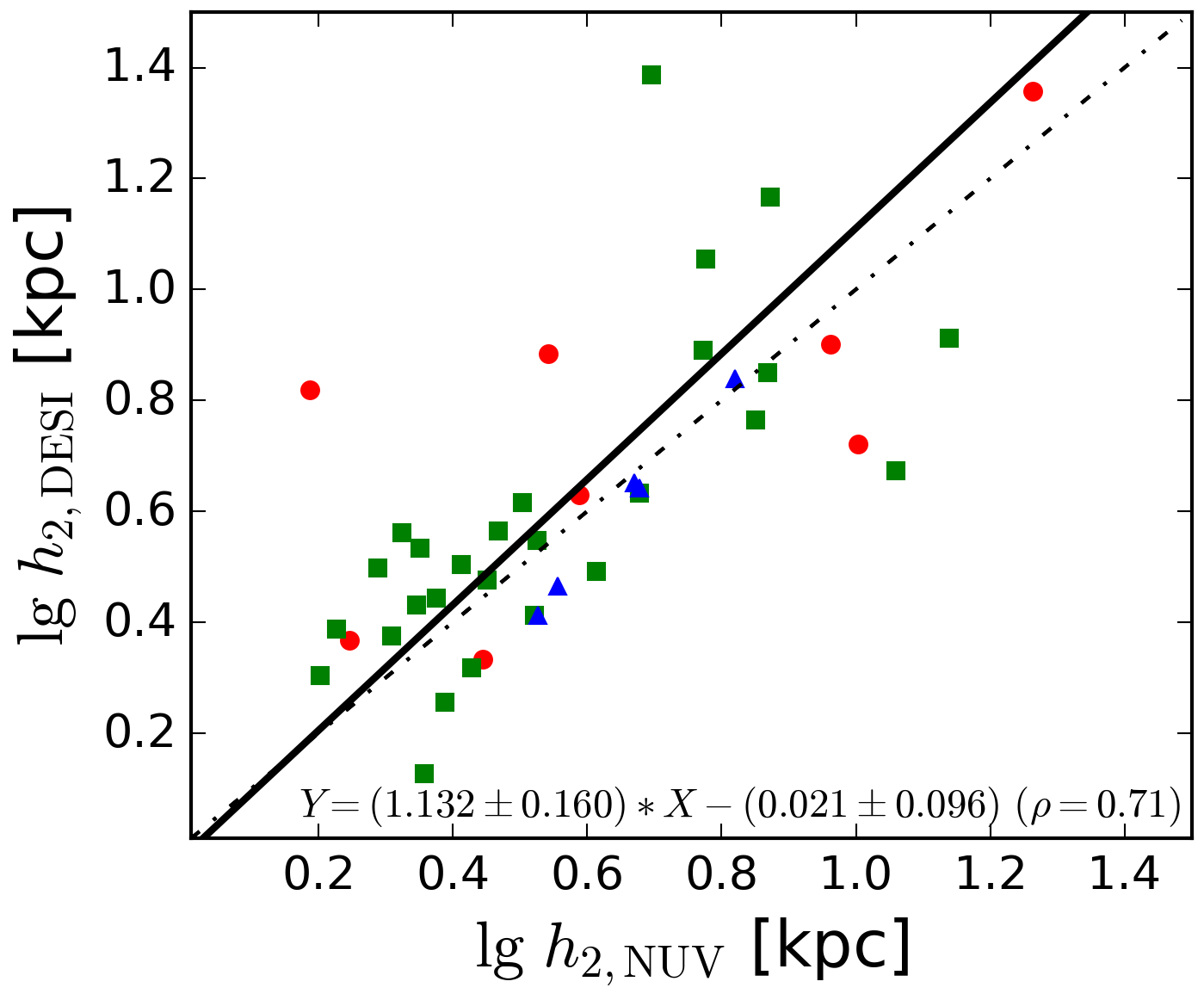}
    \caption{Comparison of the characteristics of the discs in the optical ($r$ band) and NUV: bending strength (upper left plot), break radius (upper right plot; galaxies with the same disc profiles in the NUV and $r$ band are shown by bright colors, whereas all other galaxies are depicted by pale colors), first disc scalelength (bottom left plot), and second disc scalelength (bottom right corner). For the disc scale lengths, only 47 galaxies with one- or two-disc models are shown. Linear regression equations are provided in each plot and depicted by a black solid line, whereas a one-to-one relationship is shown by a dashdot line. $G$, $M$ and $F$ denote grand-design, multiarmed and flocculent galaxies, respectively. Median values with 25th and 75th percentiles for each class are provided in the legend.}
    \label{fig:DES_vs_NUV}
\end{figure*}

In Fig.~\ref{fig:DES_vs_NUV}, we compare different characteristics of disc profiles in the NUV and $r$ bands: the bending strengths $BS$ (left upper plot), the break radii $R_\mathrm{br}$ (right upper plot -- only galaxies, having the same type of disc profile in the optical and NUV, are displayed), and the first (inner) and second disc scale lengths. As we can see, the bending strengths show a moderate correlation, with $BS_\mathrm{DESI}$ being systematically slightly higher than $BS_\mathrm{NUV}$, especially for multiarmed spirals with $BS_{NUV}<-50$ (the uncertainty of the intercept is significantly lower than the intercept value). This indicates that the DESI disc profiles are, on average, somewhat more up-bending than the corresponding NUV profiles. We can notice that in the NUV, there are a number of multiarmed galaxies with a Type I disc profile ($BS=0$), whereas the same galaxies in the optical may possess up- or down-bending disc profiles. This difference in the NUV and optical profiles cannot be explained by the difference in the resolutions of the NUV images with $BS_\mathrm{NUV}\approx0$ and $BS_\mathrm{NUV}\neq0$: $R_{25}=43.2^{+30.8}_{-11.6}$~arcsec for $BS_\mathrm{NUV}\approx0$ versus $R_{25}=48.9^{+36.7}_{-14.6}$~arcsec for galaxies with $BS_\mathrm{NUV}\neq0$. Interestingly, the grand-design galaxies in our sample are located, on average, slightly higher above the regression line than the other galaxies and do not demonstrate extremely low bending strengths. The good trend between the two BS quantities measured in the $r$ and NUV bands serves as indirect evidence that the proposed method for estimating the bending of the disc profile is robust, which makes it useful for quantifying the type of disc profile for a further uniform comparison at different wavelengths in future studies (including comparisons with cosmological hydrodynamical simulations).

The break radii for discs that have identical types of disc profile in the NUV and $r$ bands (52 galaxies in total) are in very good agreement (see the upper right plot in Fig.~\ref{fig:DES_vs_NUV}). This suggests that approximately one-third of galaxies in our sample have very similar profiles in the wavebands under study. For these galaxies, the break radii of the disc appear to be very similar, with a 1-$\sigma$ scatter of 0.015~dex from the regression line that within the uncertainty follows the one-to-one relationship. However, two-thirds of the sample galaxies have quite different disc profiles in the NUV and $r$ bands, although the break radii in these two bands still follow the same trend but with a much larger scatter (0.1~dex), especially for the M-class spirals.

Finally, the comparison of the disc scale lengths in the NUV and $r$ bands for the innermost disc region exhibits an apparent trend with a 1-$\sigma$ scatter of 0.13~dex (bottom left plot in Fig.~\ref{fig:DES_vs_NUV}), with the optical scale lengths being, on average, systematically lower than the scale lengths in the NUV. This correlation results in the $h_\mathrm{1,DESI}$/$h_\mathrm{1,NUV}=0.74$ ratio, in perfect agreement with $h_{0.7\mu m}/h_{NUV}=0.74$ for a sample of 18 nearby spiral galaxies from \citet{2017A&A...605A..18C}. The ratio $h_\mathrm{1,DESI}$/$h_\mathrm{1,NUV}=0.74$ suggests that the galaxy discs build inside-out \citep{2007ApJ...658.1006M,2015MNRAS.451.2324P,2019ApJ...884...99F}. This is usually explained by the fact that the gas inflowing at later epochs has a higher angular momentum and, therefore, forms a more extended star-forming disc \citep{1980MNRAS.193..189F,1998MNRAS.295..319M}.
Interestingly, for the outer disc, the scale lengths in the NUV and the optical result in the different ratio $h_\mathrm{2,DESI}$/$h_\mathrm{2,NUV}=1.44$ (bottom right plot in Fig.~\ref{fig:DES_vs_NUV}). Although we admit that the number of data points is insufficient for drawing a reliable conclusion, the uncertainties of the slope and the intercept of the regression line indicate that the outer discs have a similar or even larger outer disc scale length in the optical making the actual formation process more complex because star formation in galaxies is also regulated by stellar feedback, feedback from active galactic nuclei, and external factors \citep{2001MNRAS.327.1334V,2011MNRAS.410.1391A,2014MNRAS.437.1750M,2016MNRAS.458..242T}. Note, however, that these results may be somehow affected by the fact that the presence of spiral arms can slightly (insignificantly) influence the derivation of the inner disc scale length, but has more effect on the outer disc profile (see Sect.~\ref{sec:disk_dec_vs_spirals}). In addition, we have neglected the effect of dust attenuation in galaxies, which can be quite severe at NUV-blue wavelengths \citep{2010MNRAS.403.2053G}. However, Baes et al. (in prep.) recently used the TNG50 simulation \citep{2019MNRAS.490.3196P,2019MNRAS.490.3234N}, post-processed with the radiative transfer code SKIRT \citep{2015A&C.....9...20C,2020A&C....3100381C}, and demonstrated that stellar population gradients are the main factor for the dependence of the galaxy effective radius on wavelength and the differential dust attenuation plays a lesser (but still significant) role. In principle, the estimation of the scale length for the inner disc should be more affected by dust than that of the outer disc because the dust is more concentrated towards the inner galaxy region and primarily obscures the central regions of galaxies \citep[see e.g.][]{2009ApJ...701.1965M,2010A&A...518L..72P,2015A&A...581A.103G,2018A&A...616A.120M}.

\begin{figure}
    \centering
    \includegraphics[width=1.0\linewidth]{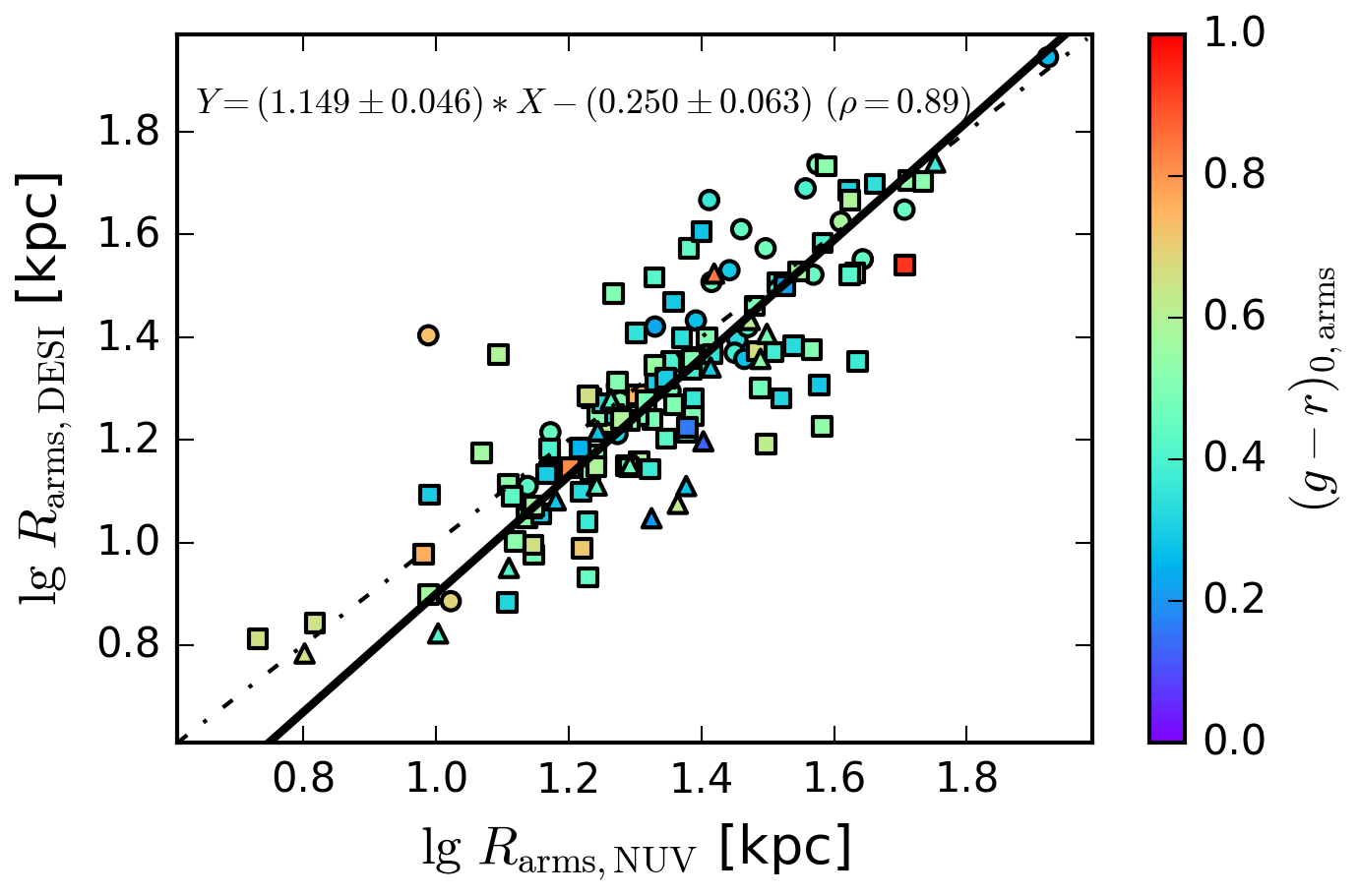}
    \caption{Comparison of the extents of spiral arms as traced in the DESI and NUV images. The $(g-r)_0$ color of the spiral arms is taken from \citetalias{Savchenko_2020} and has been corrected for Galactic extinction. The notations are the same as in Fig.~\ref{fig:DES_vs_NUV}.}
    \label{fig:DES_vs_NUV_arms}
\end{figure}

Finally, it is of great interest to compare the extent of spiral structure in the optical and NUV. From Fig.~\ref{fig:DES_vs_NUV_arms}, we can ascertain that the correlation between the two measurements is very good, close to the one-to-one relation for more extended spirals and systematically declining from it toward smaller DESI radii for less extended spirals: the effect is significant which is reflected in the small uncertainties of the slope and the intercept of the regression line. Our results suggest that the distributions of galaxies of different class of spiral structure, morphological type, spatial environment, angular size, and of different colour of spiral arms do not depend on the position on this correlation. Therefore, the deviation of the short spirals from the one-to-one relationship is not clear and requires further investigation. We speculate that it may be related to the formation history of the galaxies: larger spiral galaxies more likely experienced multiple interactions than smaller spirals, which could affect the appearance of their spiral arms in both the NUV and optical.

\subsection{Spiral structure and the types of disc profiles}
\label{sec:disk_types_vs_spirals}

Since in the previous subsection we mentioned the types of disc profiles, it is timely to discuss the statistics of the disc types in our sample and their relation to the classes of spiral structure. We also use the $BS$ indicator to quantify this comparison.

\begin{figure*}
    \centering
    \includegraphics[width=0.48\linewidth]{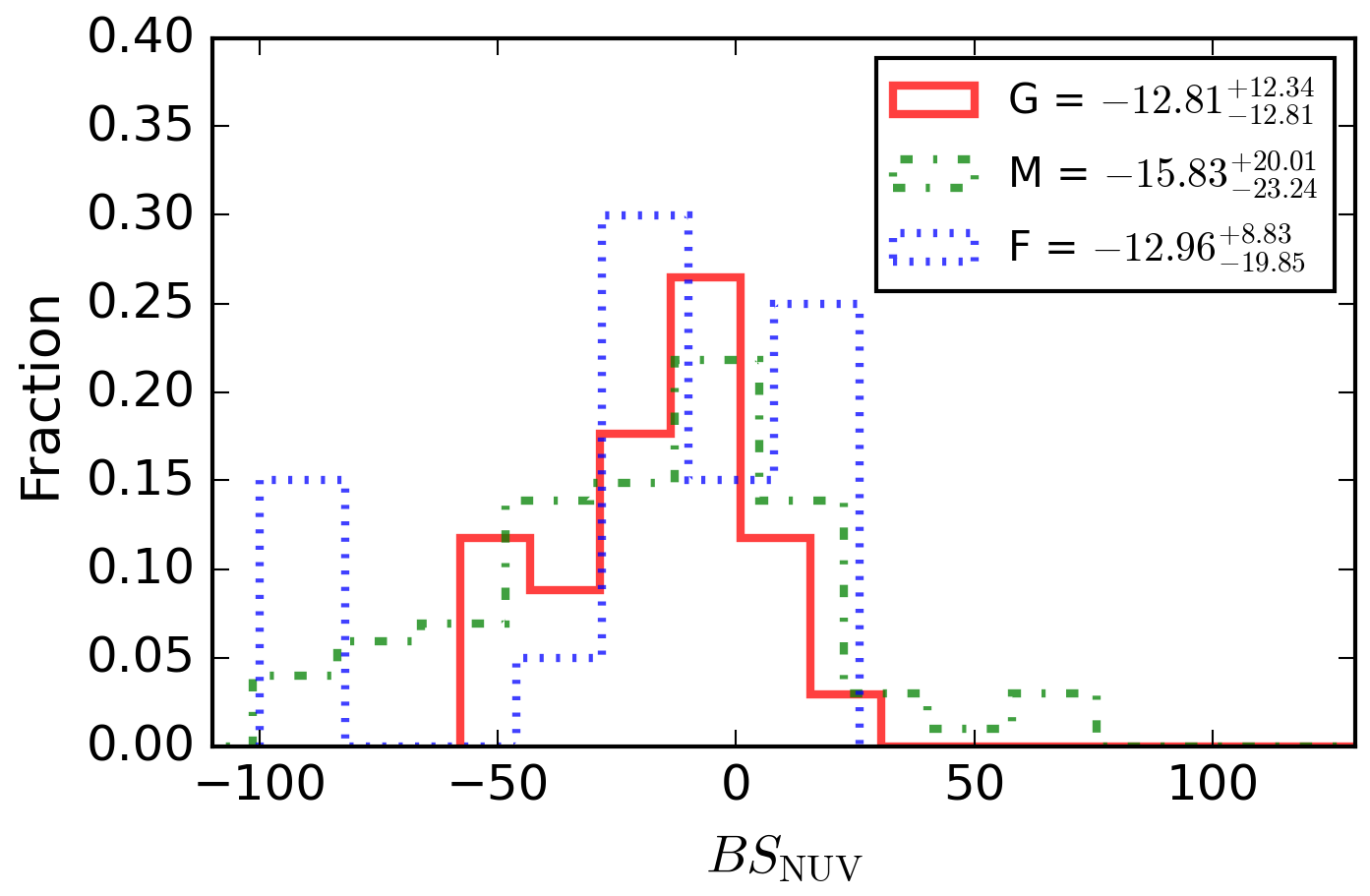}
    \includegraphics[width=0.48\linewidth]{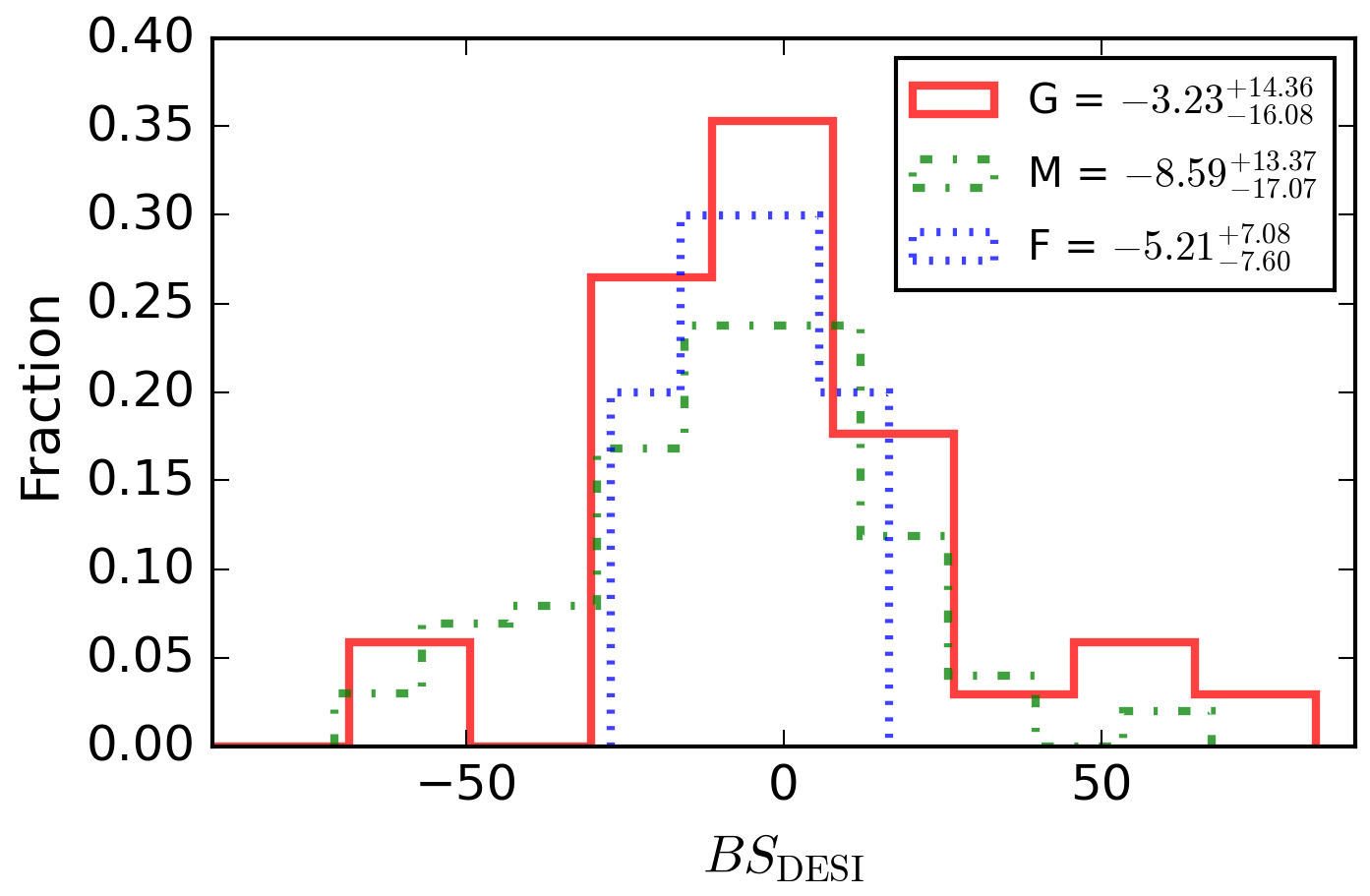}
    \caption{Distributions by bending strength indicator in the NUV and DESI $r$ band for our sample galaxies.}
    \label{fig:BS_hists}
\end{figure*}

\begin{figure*}
    \centering
    \includegraphics[width=0.4\linewidth]{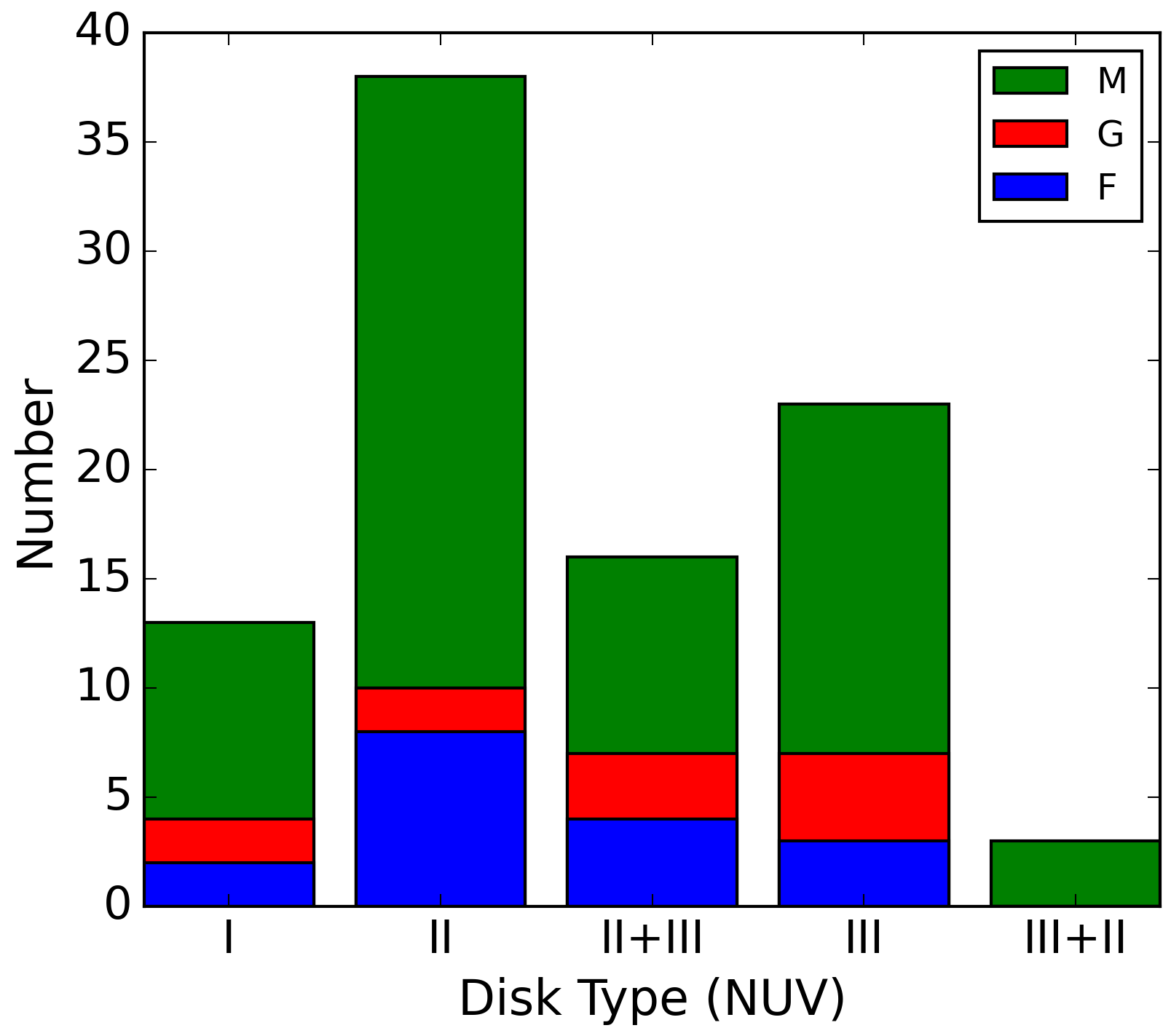}
    \includegraphics[width=0.4\linewidth]{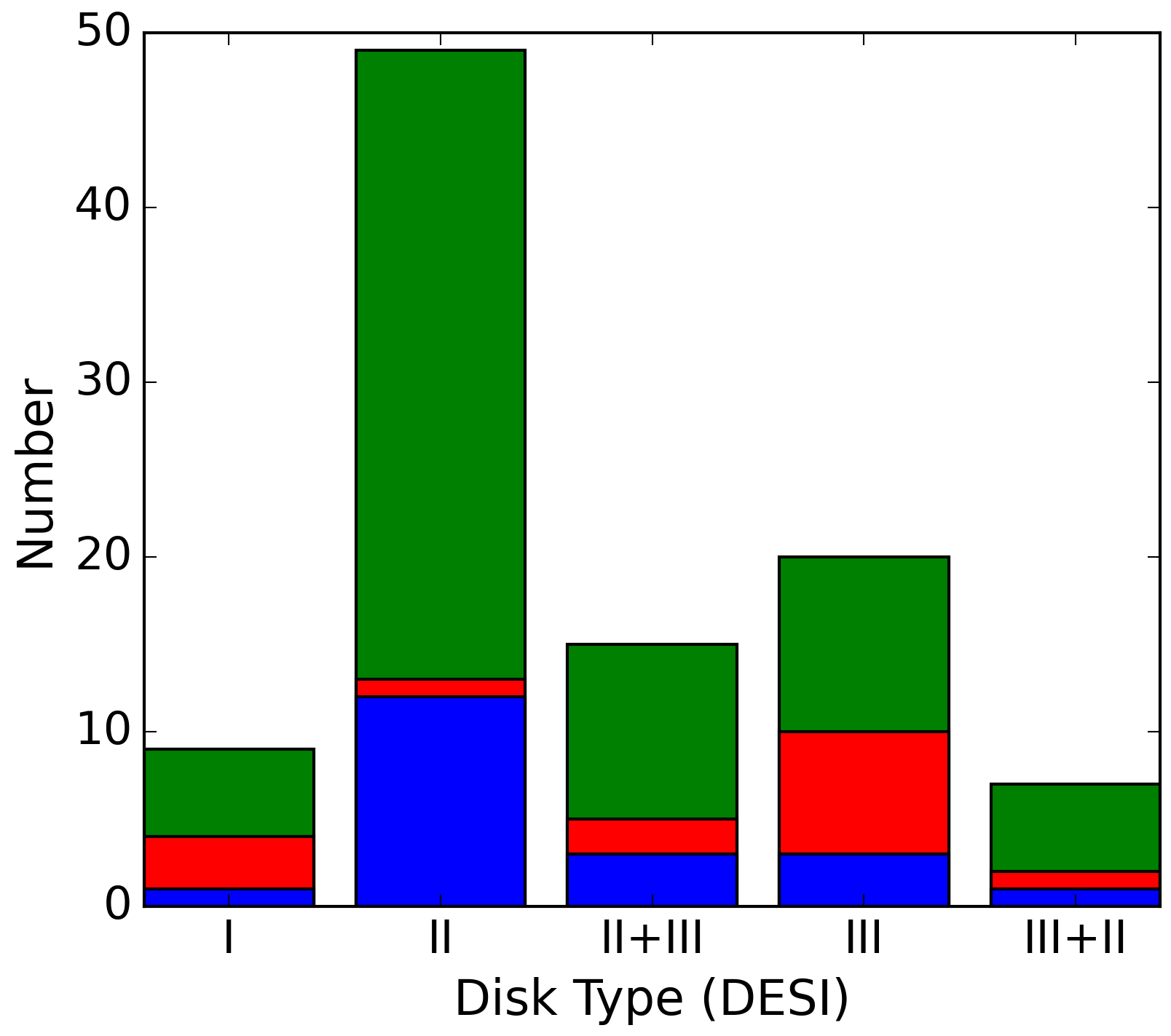}
    \caption{Distributions by disc type, determined for the NUV and DESI galaxy profiles, for the three classes to spiral structure.}
    \label{fig:profile_types}
\end{figure*}

In Fig.~\ref{fig:BS_hists}, we present distributions of the $BS$ indicator for the three classes of spiral structure in both NUV and optical data. In these histograms, we do not see any strong correlation between the type of disc profile and the class of spiral structure: grand-design, multiarmed, and flocculent galaxies are not significantly different in the $BS$ distribution. We can notice, however, that in the optical the distributions appear almost symmetric and only slightly shifted toward negative values, whereas in the NUV the distributions are more dominated by negative bending strengths with overall down-bending profiles. This fact is also demonstrated in Table~\ref{tab:BS_types} where one can see, inter alia, the robustness of the $BS$ indicator in quantifying the shape of disc profiles. The column heights in Fig.~\ref{fig:profile_types} confirm that down-bending profiles are the most common type in all three classes of spiral structure, although grand-design galaxies have a higher fraction of Type III and Type II+III discs than the other two classes. Flocculent galaxies in the optical are significantly dominated by Type II profiles and have a tiny fraction of single-exponential discs. In general, the Type I profile is the least common type in our sample for all types of spirals (there is also a small fraction of Type III+II among multiarmed spirals). Although quantitatively the distributions by the disc type in the NUV and optical do not quite match (the fraction of Type II discs is appreciably higher in DESI, whereas in GALEX the fractions of other types are higher relative to the down-bending discs), the overall impression remains the same as we have discussed above.

Our results agree well with the literature: \citet{2006A&A...454..759P} determined that only 10\% of their sample of late-type spirals belong to Type I discs, 60\% have Type II profiles, and 30\% demonstrate up-bending profiles. The distribution by types changes for early-type barred disc galaxies \citep{2008AJ....135...20E}: 27\% of Type~I, 42\% of Type~II, 24\% of Type~III, and 6\% for combinations of Type II and Type III. \citet{2011AJ....142..145G} studied the surface brightness profiles of early-type, unbarred disc galaxies and, based on previous studies by \citet{2006A&A...454..759P} and \citet{2008AJ....135...20E}, concluded that Type I profiles have a higher fraction in early-type disc galaxies and are least frequent in Sd-Sm spirals. In contrast, the frequency of Type II profiles increases from lenticular to the latest-type spirals.
They also provide the fractions of disc profile types for S0-Sm disc galaxies: 21\% (Type I), 50\% (Type II), 38\% (Type II) including 8\% with composite Type II+III profiles. These frequencies agree well with our results for DESI observations listed in Table~\ref{tab:BS_types} taking into account that our sample comprises a rather small fraction of early-type spirals. Note that in the mentioned studies, the authors considered surface brightness profiles down to levels below 27-28~mag\,arcsec$^{-2}$ in the $R$ band that are similar to the surface brightness limits that we set for our galaxy profiles.

\begin{table}
\caption{Bending strength as measured for different types of disc profile. Median values with 25th and 75th percentiles are listed.}
\label{tab:BS_types}
\centering
    \begin{tabular}{lcccc}
    \hline
    \hline    
    Type &  $BS_\mathrm{NUV}$ & $BS_\mathrm{DESI}$  & \%$_\mathrm{NUV}$ & \%$_\mathrm{DESI}$ \\[0.5ex]
    \hline
    I      &   0                          &    0      &    14      &      9    \\[+0.8ex]
    II     & $-27.06^{+11.74}_{-19.33}$ & $-17.21^{+7.05}_{-8.98}$&   40       &   48          \\[+0.8ex]
    II+III & $-32.74^{-13.62}_{+22.96}$ & $-12.46^{-6.93}_{+9.38}$&    24      &    15         \\[+0.5ex]
    III    & $8.77^{-3.77}_{+11.15}$     & $14.88^{-4.38}_{+10.14}$&    20      &    22         \\[+0.8ex]
    III+II & $40.95^{-16.22}_{+35.08}$   &   $15.34^{-7.67}_{+8.33}$       &    2      & 6         \\[+0.8ex]
    \hline\\
    \end{tabular}    
\end{table}

\subsection{The outer spiral structure}
\label{sec:outer_structure}

Let us now turn to the main subject of this paper, the properties of spiral structure beyond the optical radius, as measured using the NUV and deep optical data. 

For different arm classes and spatial environments, we did not find dramatic differences within and beyond the optical radius for the following characteristics of spiral pattern: the average pitch angle, asymmetry of the spiral arms, and pitch angle variations along the radius. For example, the average pitch angle of spiral arms beyond the optical radius may be several times greater or lower than the average pitch angle within the optical radius (half of the galaxies demonstrate an increase of the pitch angle outside of the optical radius, while the other half have lower pitch angles). Interestingly, this result holds for both isolated galaxies and galaxies in groups and clusters.
Therefore, we can conclude that the pitch angle of galaxies may become systematically greater or lower in the galaxy periphery without any dependence on class of spiral structure, type of disc profile, global galaxy morphology, or spatial environment.

\begin{figure*}
    \centering
    \includegraphics[width=0.33\linewidth]{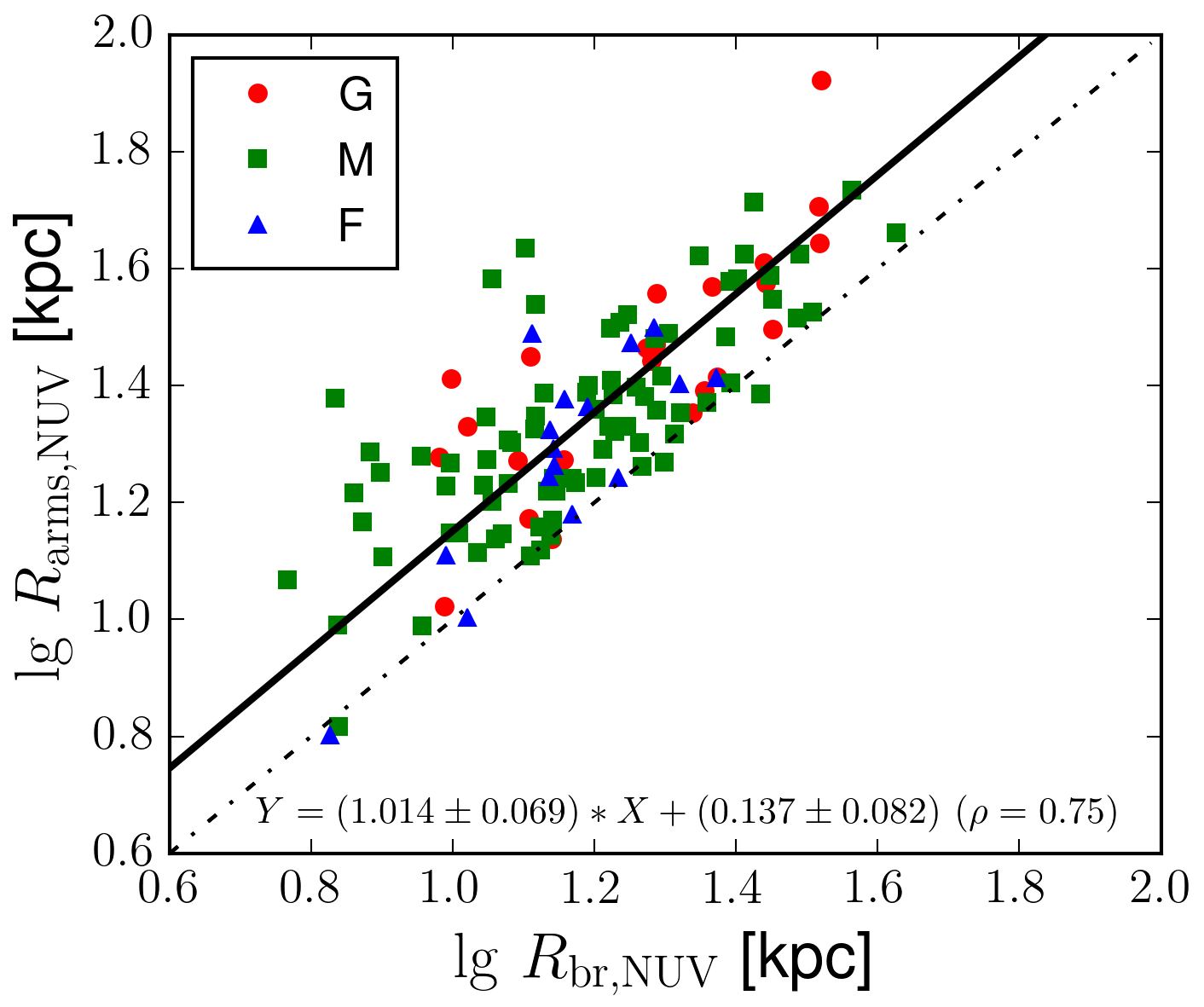}
    \includegraphics[width=0.33\linewidth]{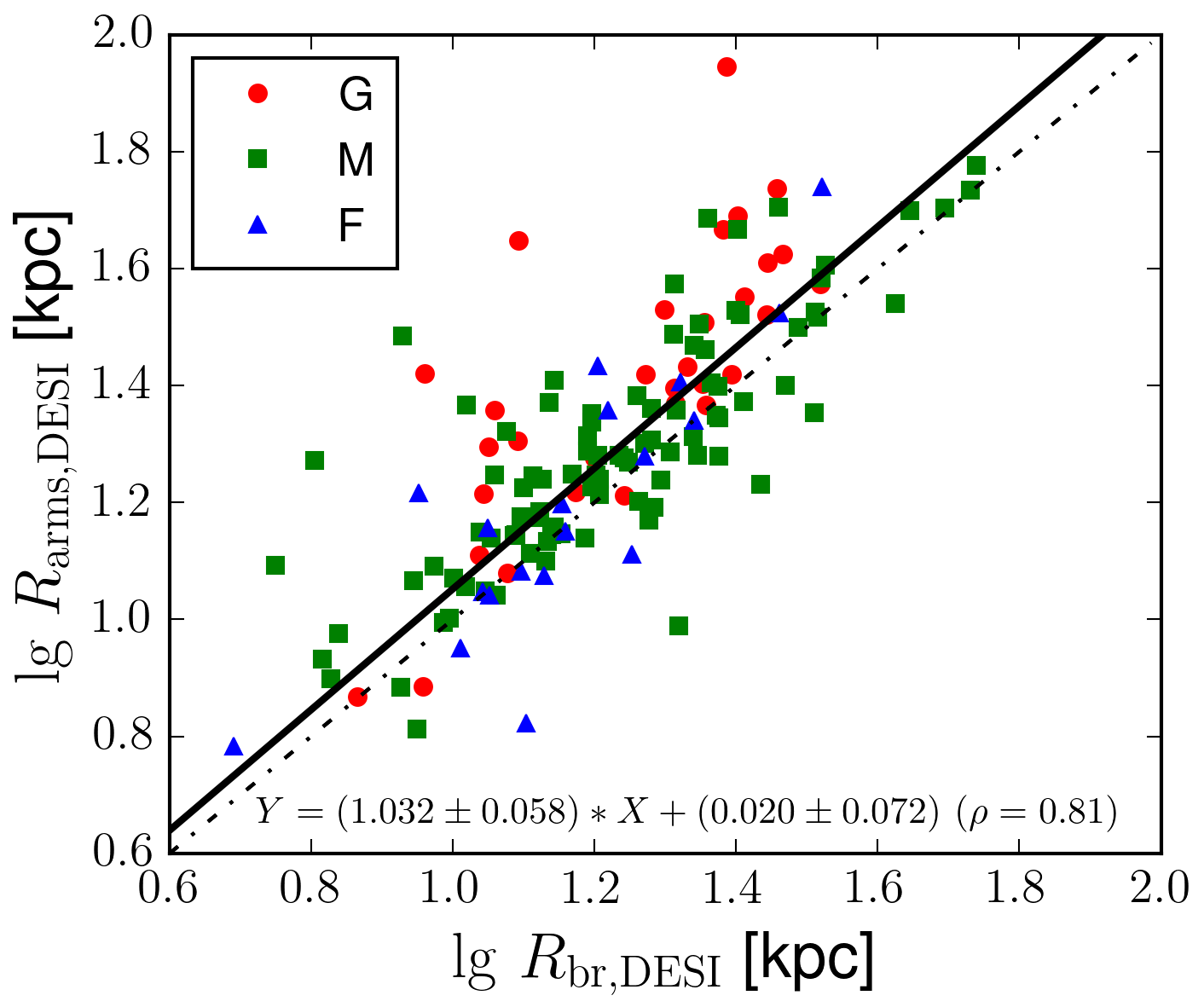}
    \includegraphics[width=0.33\linewidth]{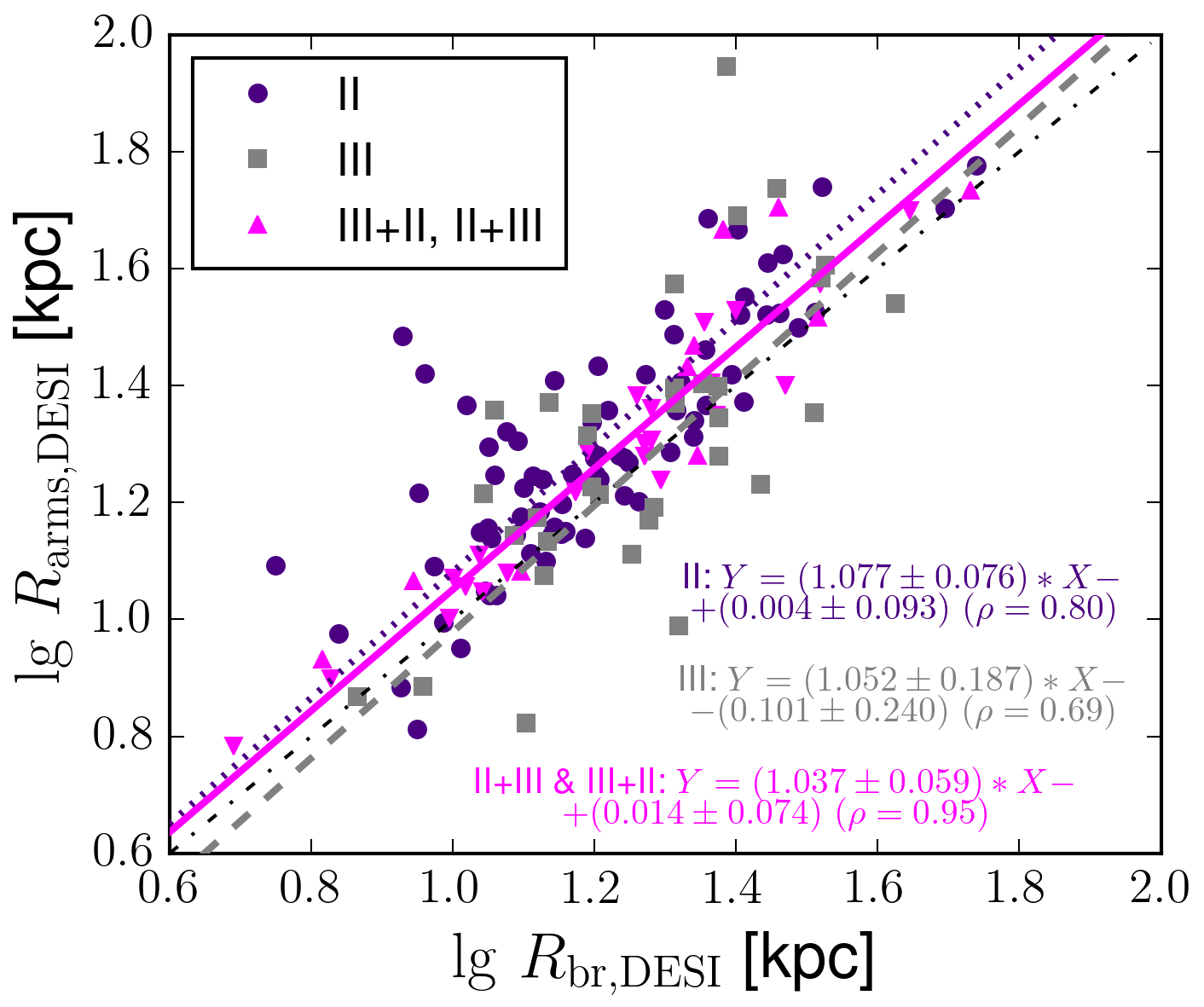}
    \caption{Correlations between the radius of spiral structure (as measured in the NUV and optical --- lefthand and middle plots, respectively) and the nearest break radius to the end of spiral structure. The right-hand plot depicts the same data points as in the middle one but with different disc types highlighted. Dot-dashed lines depict the one-to-one ratio, whereas all other lines represent regression lines for different scatter plots.}
    \label{fig:Rarms_Rbr}
\end{figure*}

\begin{figure}
    \centering
    \includegraphics[width=0.95\linewidth]{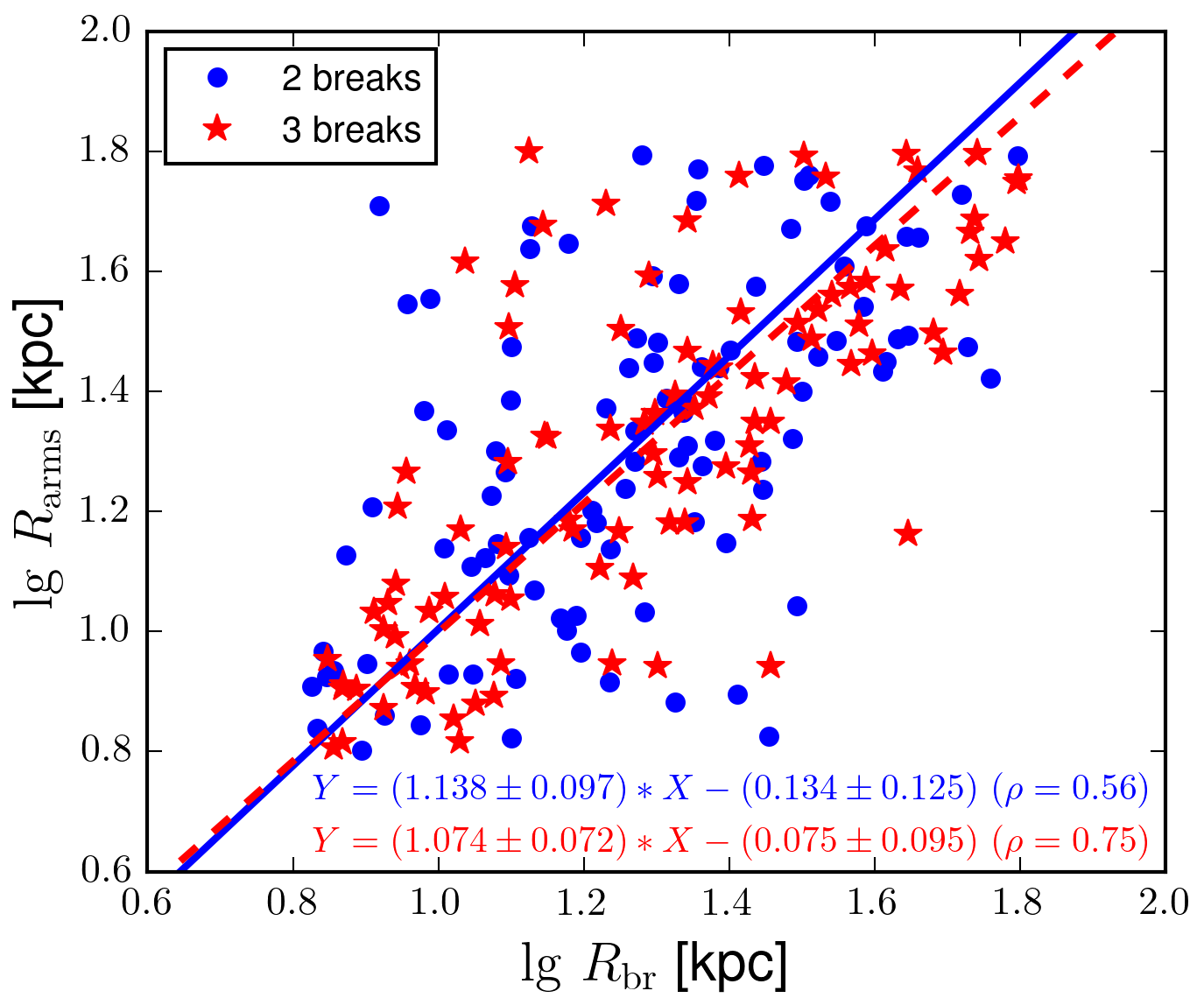}
    \caption{``Artificial'' dependence between the radius of spiral arms and the nearest break radius. Both break radii (two or three) and radius of spiral arms were generated randomly in the range [0.8,1.8]. As one can see, the larger number of breaks increases the tightness of the correlation. However, in Fig. \ref{fig:Rarms_Rbr}, righthand plot, the Pearson coefficient for double- or triple-break profiles is 0.95, whereas in these simulations (if we repeat random generation of these arrays many times), it does not exceed $\sim0.8$ for triple-break profiles.}
    \label{fig:sim}
\end{figure}

In Fig.~\ref{fig:Rarms_Rbr}, one can see a correlation between the extent of spiral arms and the nearest break radius to the edge of the spiral arms in the NUV (left plot) and DESI (middle plot) observations. Both correlations are tight (albeit with some outliers) which suggests that the extent of the spiral arms is always comparatively close to one of the disc breaks: $R_\mathrm{br,NUV} \simeq 0.73\,R_\mathrm{arms,NUV}$, $R_\mathrm{br,DESI} \simeq 0.92\,R_\mathrm{arms,DESI}$. In the NUV, the nearest break almost always lies within the radius of spiral structure, so $0.53\,R_\mathrm{arms,NUV} \lesssim R_\mathrm{br,NUV}\lesssim 1.02\,R_\mathrm{arms,NUV}$. This general trend reflects the fact that in more extended discs (and hence, for greater break radii), the spiral arms continue farther out than in galaxies with a smaller linear size. Therefore, this correlation represents a typical galaxy scaling relation: in galaxies with a larger physical size, all main characteristics scale relative to smaller galaxies. 
Note the systematic upward offset of the correlation for the NUV data with respect to the DESI data, which serves as another evidence that, on average, NUV spirals are typically more extended than optical ones. In the right plot of Fig.~\ref{fig:Rarms_Rbr}, we display the same scatter plot as in the middle one, but highlight different disc types. We can notice that galaxies with several breaks (Type II+III and Type III+II) lie closer to the general regression line, whereas discs with only one break (Type II and Type III) demonstrate a much larger scatter along the trend. Furthermore, the spiral arms in Type II discs are, on average, slightly more extended than those in Type III discs for the same break radius, although the intercepts of the respective regression lines 
overlap within their uncertainties. Specifically, the spiral arms embedded in Type II discs end, on average, at $0.95\,R_\mathrm{br,DESI} \lesssim R_\mathrm{arms,DESI} \lesssim1.29\,R_\mathrm{br,DESI}$, whereas Type III discs extend to $0.86\,R_\mathrm{br,DESI} \lesssim R_\mathrm{arms,DESI} \lesssim 1.17\,R_\mathrm{br,DESI}$. In galaxies with several disc breaks, the nearest break radii are located even closer to the radius of spiral structure: $1.01\,R_\mathrm{br,DESI} \lesssim R_\mathrm{arms,DESI} \lesssim 1.15\,R_\mathrm{br,DESI}$. However, note that when the number of disc breaks increases, the tightness of the correlation between $R_\mathrm{arms}$ and $R_\mathrm{br}$ also increases, even if these breaks are orientated randomly along the disc. In Fig.~\ref{fig:sim}, one can see, however, that this ``artificial'' effect cannot reproduce the very tight correlation we observe for galaxies with two to three breaks (note that only six galaxies in our sample demonstrate three profile breaks and 28 galaxies have two breaks). This plot provides visual evidence that at least some breaks in disc profiles should be closely related to the truncation of spiral arms.

Using Spitzer 3.6~$\mu$m observations, \citet{2014MNRAS.441.1992L} explored how the environmental density of galaxies and the tidal interaction strength can affect the shape of the disc profile. They found that most Type II profiles are observed in galaxies with apparent rings, lenses, and spiral arms. Similarly to \citet{2006A&A...454..759P}, they did not find a notable correlation between the galaxy environment and the type of disc profile, although they note a positive correlation between the disc scale length and the tidal interaction strength for Type III profiles. 

\citet{2017MNRAS.467.2127P} estimated the frequency rates of Type I, Type II and Type III profiles in field and cluster environments and reported that the abundance of Type I profiles in clusters is higher ($15_{-4}^{+7}$\%) than in the field ($6\pm2$\%) which may be explained by environment-driven mechanisms: in clusters, Type II discs are converted to Type I discs over time. 

\citet{2019A&A...625A..36W} investigated Spitzer 3.6~$\mu$m surface brightness profiles of disc galaxies, in particular, Type III disc break-hosting galaxies. They showed that most Type III breaks are caused by morphological asymmetries of the host galaxies, which, in its turn, is the result of tidal disturbance because these galaxies reside in dense environments. \citet{2019MNRAS.486.1995S} also note that some of the surface brightness profile breaks may be due to the presence of stellar haloes. \citet{2019A&A...625A..36W} point out that Type III disks are often associated with outer spheroidal components as a result of, for example, galaxy harassment. Finally, some Type III breaks, residing in low density environments, are related to extended spiral arms or star formation (secular evolution features), and in such galaxies down-bending breaks are often found at larger radii. These conclusions are confirmed in a recent study by \citet{2022MNRAS.509..261P} who used EAGLE \citep{2015MNRAS.446..521S} cosmological hydrodynamical simulations to explore the effects of galaxy mass and spatial environment on types of galaxy profiles. Down-bending breaks highlight truncations of star-forming discs, which, in their turn, imply radial gradients in the galaxy stellar populations. Consistent with observations, the frequency of Type II discs in simulations increases with Hubble stage. Type III profiles may form through several scenarios, the main of which is galaxy mergers. Also, in some Type III galaxies a shorter inner disc may have formed at later cosmological epochs. It comes as no surprise why the frequency of Type III galaxies increases with galaxy mass, since for more massive galaxies, galaxy mergers have been more frequent. The high fraction of Type I galaxies in galaxy clusters is explained by star formation quenching, which takes place when a star forming Type II galaxy enters a galaxy cluster and its young stellar population fades away converting it to a typical Type I galaxy.

To investigate how the current shape of our sample galaxies may be affected by tidal interactions, we examined their deep DESI images (our classification is provided in Table~\ref{tab:spirals}). We found that only a tiny fraction of our spiral galaxies (12 out of 155 galaxies, or $\approx8$\% of the sample) demonstrate signs of low-surface brightness structures brighter than 27~mag\,arcsec$^{-2}$ in the $r$ band (loops, arcs, bridges, stellar streams, etc.). This subsample comprises 7 multiarmed galaxies and 5 grand-design spirals, and no flocculent galaxies. Also, in accordance with \citet{2023A&A...671A.141M}, some galaxies in our sample (for example, PGC\,22957, PGC\,24996, PGC\,36902) exhibit galactic feathers produced by interactions with low-mass companions that perturb the disc of the host and lead to the formation of narrow tidal tails resembling the outer structure of asymmetric spiral arms in some galaxies from our sample. A similar mechanism involves interactions with a more massive neighbour that may induce the formation of grand-design spiral arms in the interacting galaxy \citep[see e.g.][]{1941ApJ....94..385H,1972ApJ...178..623T, 1973A&A....22...41E,2016AJ....152..150G}. In this scenario, tidal-bridge tails are induced by fly-bys of nearby companions, which later wind-up and transform into grand-design spirals. However, these structures dissipate within 1 Gyr due to the decrease of their pitch angle with time, as they continue to wind up \citep{2008ApJ...683...94O,2015ApJ...807...73O}. The formation of grand-design spiral arms is also possible in the potential of a galaxy cluster, without any perturbations from close companions \citep{2017ApJ...834....7S}.

\begin{figure*}
    \centering
    \includegraphics[width=0.33\linewidth]{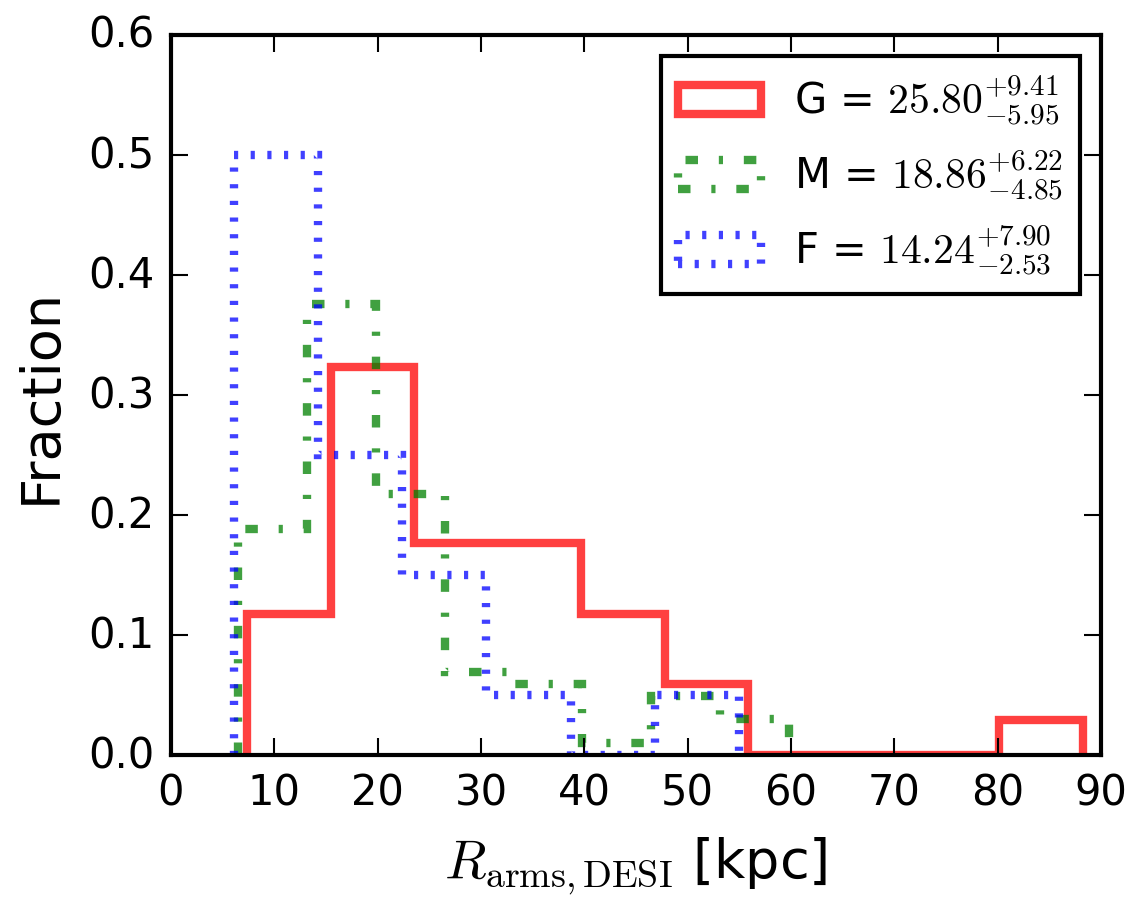}
    \includegraphics[width=0.33\linewidth]{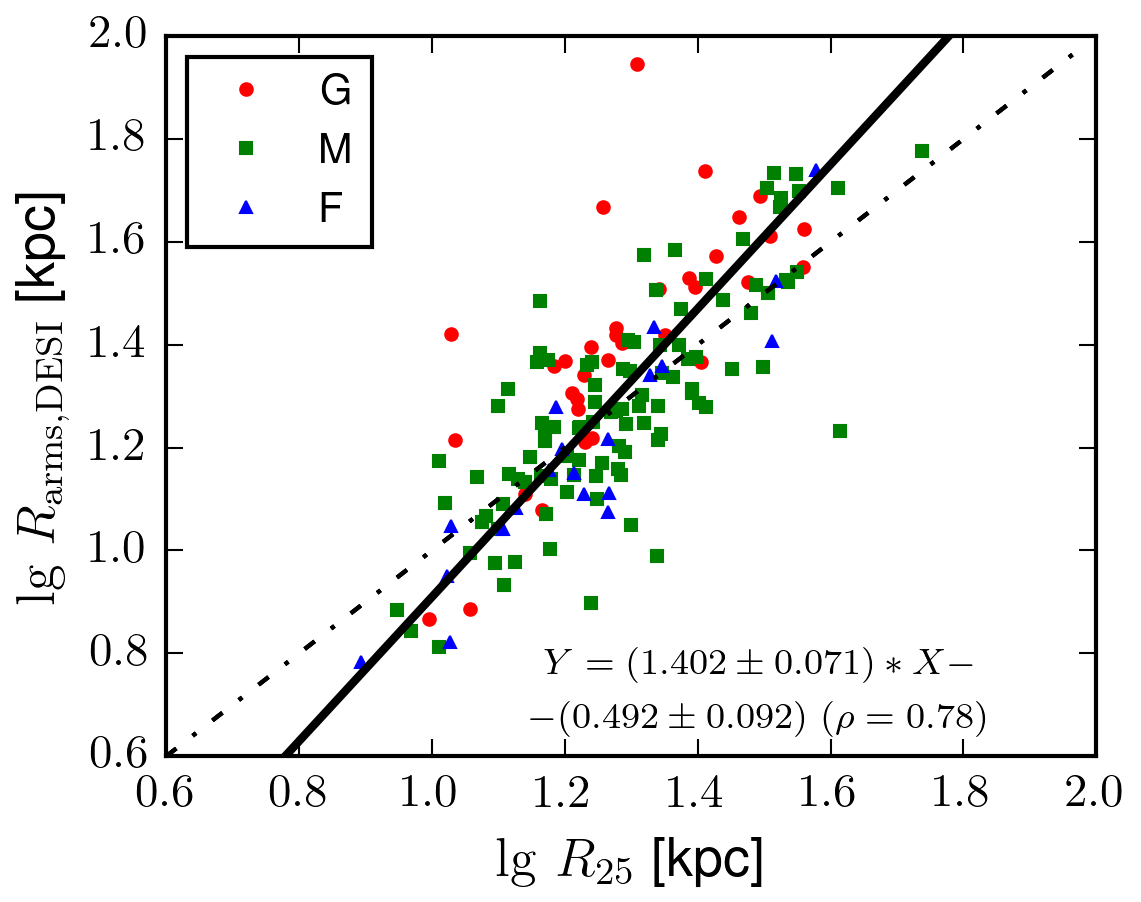}
    \includegraphics[width = 0.33\linewidth]{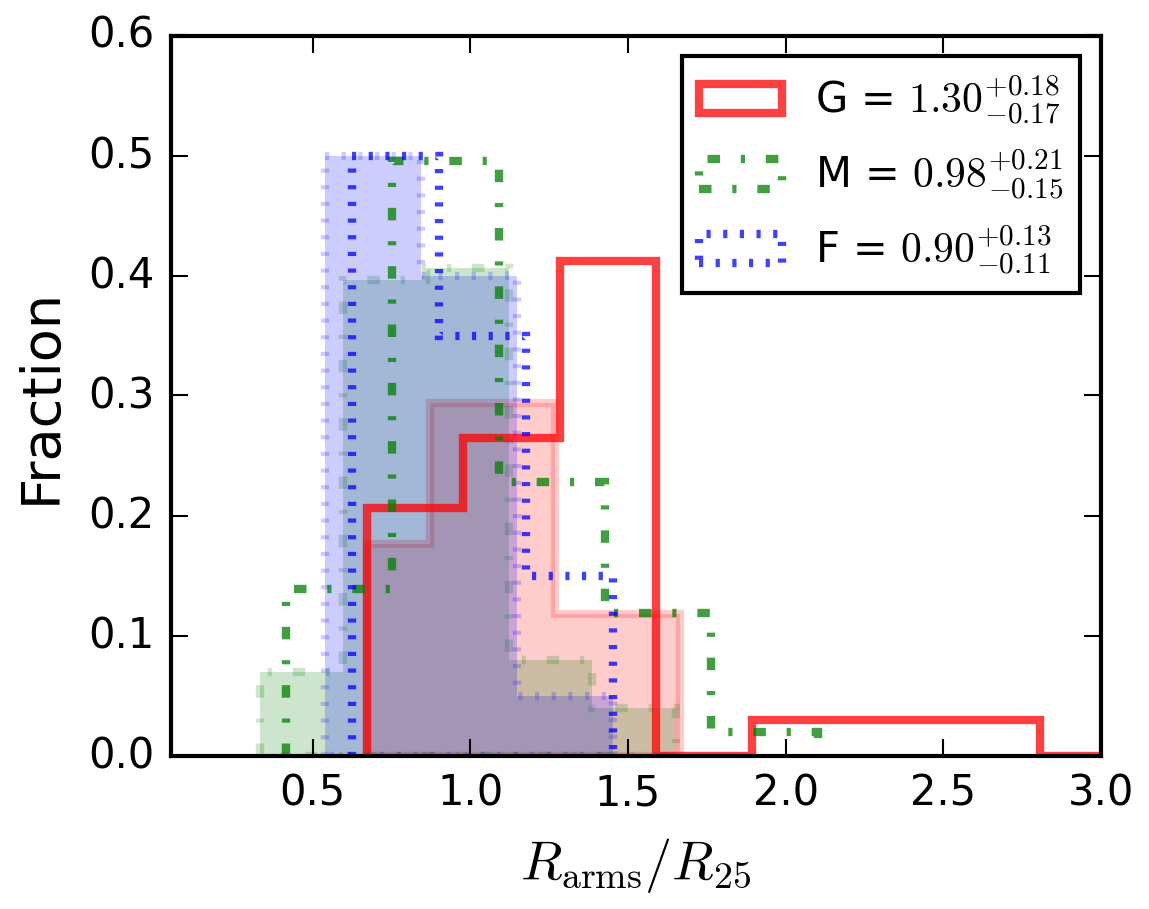}
    \caption{Distributions of the radial extent of spiral arms: left plot for absolute units and right plot for units expressed in optical radii. The middle plot shows a correlation between the radial extent of spiral arms and the optical radius. In the right plot, the coloured histograms show results of \citetalias{Savchenko_2020} obtained with use of regular SDSS imaging, whereas the unfilled histograms depict results of this paper. Note that the vast majority of grand-design galaxies lie above the general trend (the solid line in the middle plot) and are better described by $Y = (1.500\pm0.184)\,X - (0.526\pm0.238)$ with $\rho=0.73$, although they are still located within the scatter of the general trend.}
    \label{fig:arms_extent}
\end{figure*}

In Fig.~\ref{fig:arms_extent}, we consider the extent of spiral arms in the sample galaxies. In the left plot, we can see that grand-design galaxies are, on average, more extended than flocculent and multiarmed galaxies. We can also conclude that the maximum extent of the spiral arms in our sample for all three classes reaches 55--60~kpc, with an exception of PGC\,22957 with $R_\mathrm{arms,DESI}=88.3$~kpc. In the middle plot of Fig.~\ref{fig:arms_extent}, one can see a correlation between the optical radius and the extent of spiral arms. For most galaxies, the deviation from the one-to-one relation does not exceed 0.15~dex which translates into $0.71\,R_{25}\lesssim R_\mathrm{arms,DESI} \lesssim 1.41\,R_{25}$. Interestingly, grand-design galaxies typically lie above the general trend -- their spiral arms extend beyond the optical radius. This is also illustrated in the right plot of Fig.~\ref{fig:arms_extent}, where the extent of spiral arms is normalised by the optical radius of each galaxy. While the other two classes of spiral structure peak at $R_\mathrm{arms,DESI}\simeq(1.0\pm0.2)\,R_{25}$, the distribution of grand-design galaxies by this parameter is asymmetric and shifted towards $R_\mathrm{arms,DESI}\simeq1.4\,R_{25}$. This suggests that grand-design galaxies are more likely to have been stretched by past tidal interactions with neighbouring galaxies than the other two classes of spirals. In the same right plot of Fig.~\ref{fig:arms_extent}, we show the maximum extent of the spirals as measured by \citetalias{Savchenko_2020} using SDSS observations (colour-filled histograms). As one can see, the increase of the arms' radius in deep images $\langle R_\mathrm{arms,DESI}/R_\mathrm{arms,SDSS}\rangle$ is moderate for flocculent  ($1.12\pm0.16$) and multiarmed ($1.18\pm0.24$) galaxies, whereas for G-arms it is more appreciable ($1.28\pm0.36$). Therefore, deep photometry is especially important for studying the extent of grand-design galaxies, whereas most flocculent and multiarmed spirals can be almost entirely traced in ordinary SDSS observations.

\begin{figure}
    \centering
    \includegraphics[width = 0.9\linewidth]{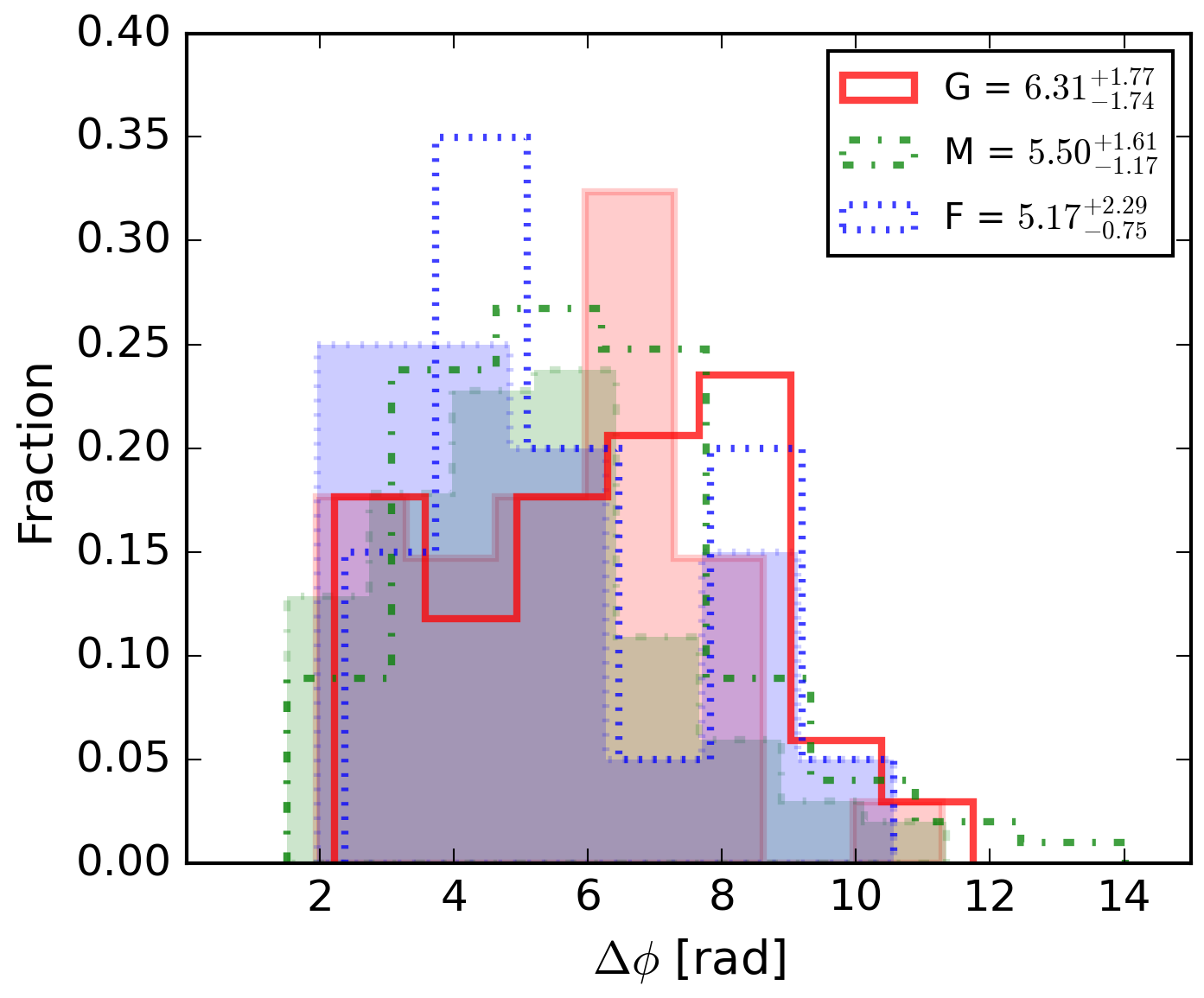}
    \caption{Distribution by the azimuthal angle for the sample galaxies. The coloured histograms show results of \citetalias{Savchenko_2020} obtained with use of regular SDSS imaging, whereas unfilled histograms display results of this paper.}
    \label{fig:phimaxl}
\end{figure}

Fig.~\ref{fig:phimaxl} shows a distribution by the overall azimuthal angle of spiral arms in our sample, separated by class of spiral structure (grand design, multiarmed, and flocculent). The data demonstrate that spiral arms in grand-design galaxies exhibit slightly larger maximum azimuthal angles, on average, than the other two classes, whereas flocculent galaxies typically have less than one full turn of spiral arms. As in the case with the maximum radius of spiral arms, the advantage of using deep photometry for studying spiral structure is evident for grand-design ($\Delta \phi_\mathrm{DESI}/\Delta \phi_\mathrm{SDSS}=1.18\pm0.24$) and multiarmed spirals ($1.18\pm0.27$), while flocculent galaxies show a slightly lower increase ($1.13\pm0.19$).

\begin{figure*}
    \centering
    \includegraphics[width = 0.48\linewidth]{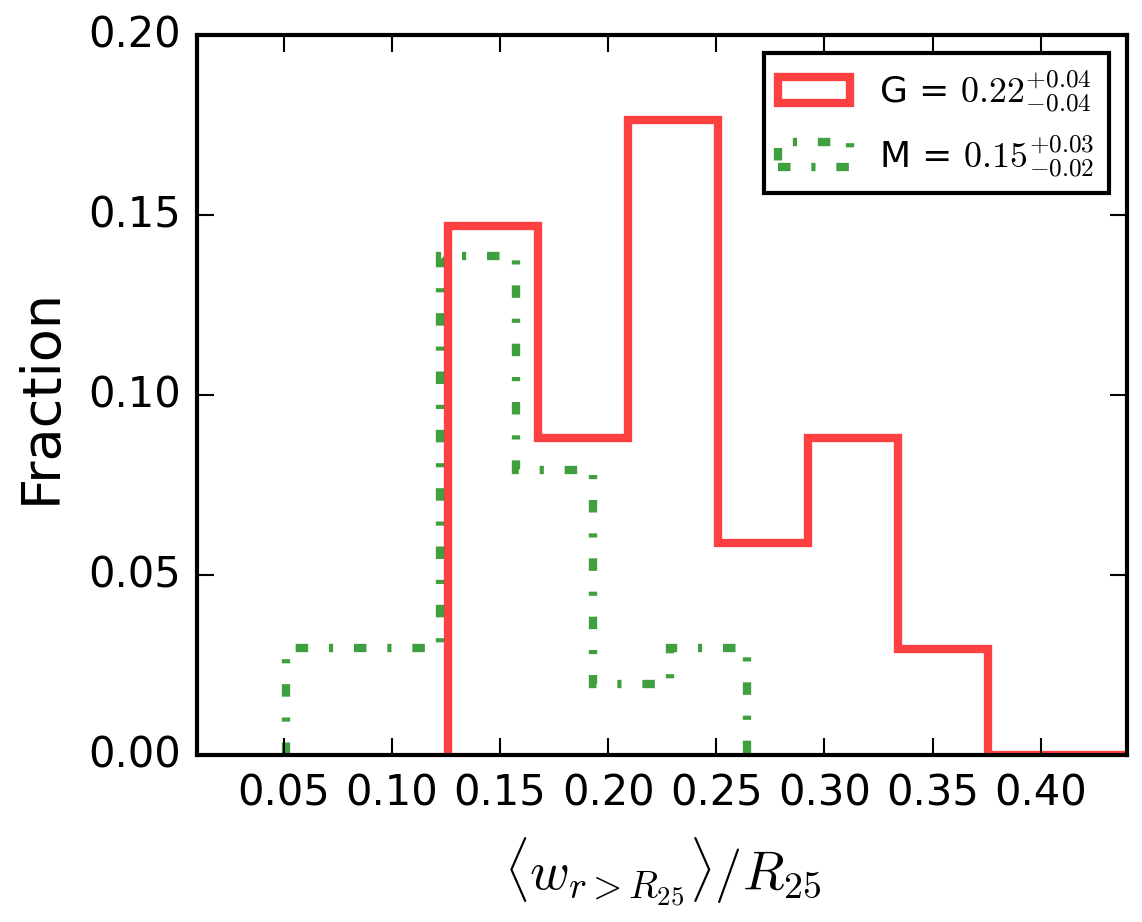}
    \includegraphics[width = 0.48\linewidth]{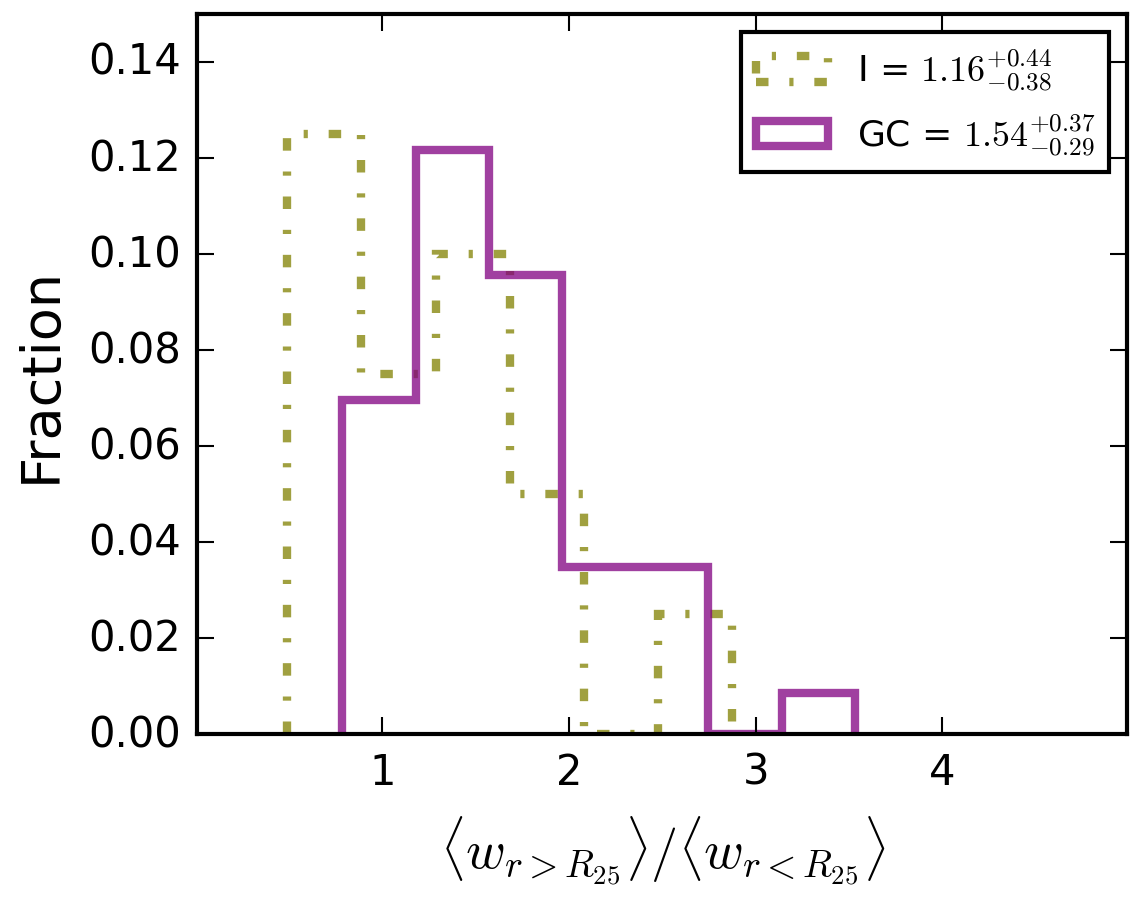}
    \caption{Left plot: distribution of the average spiral arm width outside the optical radius and normalised by the optical radius. Flocculent galaxies are not shown because only few spirals of this class have radii larger than the optical radius. Right plot: 
  distributions of isolated galaxies (I) and galaxies in groups and clusters (GC) by the average spiral arm width outside the optical radius, normalised by the average spiral arm width inside the optical radius taken from \citetalias{Savchenko_2020}.}
    \label{fig:widthall}
\end{figure*}

The measurements of arm width, taken at each point beyond the optical radius, were averaged to obtain a single arm width value (see the distribution of $\langle w \rangle/R_{25}$ in Fig.~\ref{fig:widthall}, left plot). Although the arm width distributions, obtained for SDSS images alone, do not display much scatter (see figure~16 in \citetalias{Savchenko_2020}), the deeper DESI images show that grand-design galaxies have, on average, wider spiral arms than the other two classes of spirals. 
Tidal interactions could also explain this increase in arm width beyond the optical radius. Fig.~\ref{fig:widthall}, right plot, compares the measured width of spiral arms beyond the optical radius for isolated galaxies versus galaxies in groups and clusters. The outer arm widths beyond the optical radius are normalised by the arm width within the optical radius. The data show that isolated galaxies tend to have smaller spiral arm widths, probably because isolated galaxies do not experience tidal effects. Interestingly, we did not find a significant difference between the extent of spiral arms (normalised by the optical radius) and the spatial environment, perhaps because our sample comprises only a small fraction of truly interacting galaxies with tidally induced arms. 

In view of the above discussion, we conclude that there is a link between the spiral structure of grand-design galaxies and interactions with neighbouring galaxies. This suggests that tidal interactions can be the main formation mechanism for grand-design spirals, while the spiral arms in multiarmed and flocculent galaxies may be predominantly formed by other mechanisms.

\subsection{The influence of spiral arms on the retrieved parameters of the disc}
\label{sec:disk_dec_vs_spirals}

\begin{figure*}
    \centering
    \includegraphics[width = 0.4\linewidth]{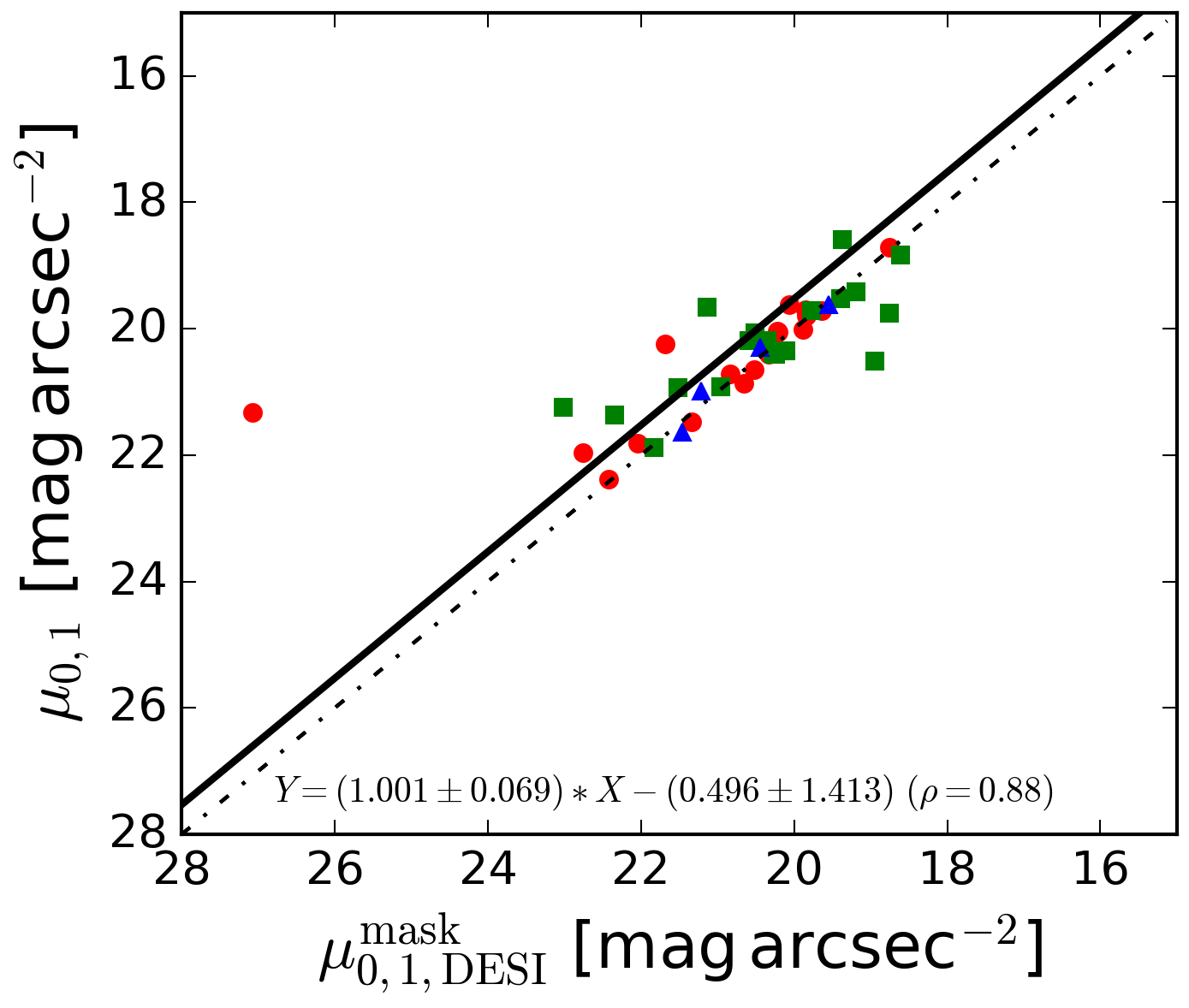}
    \includegraphics[width = 0.4\linewidth]{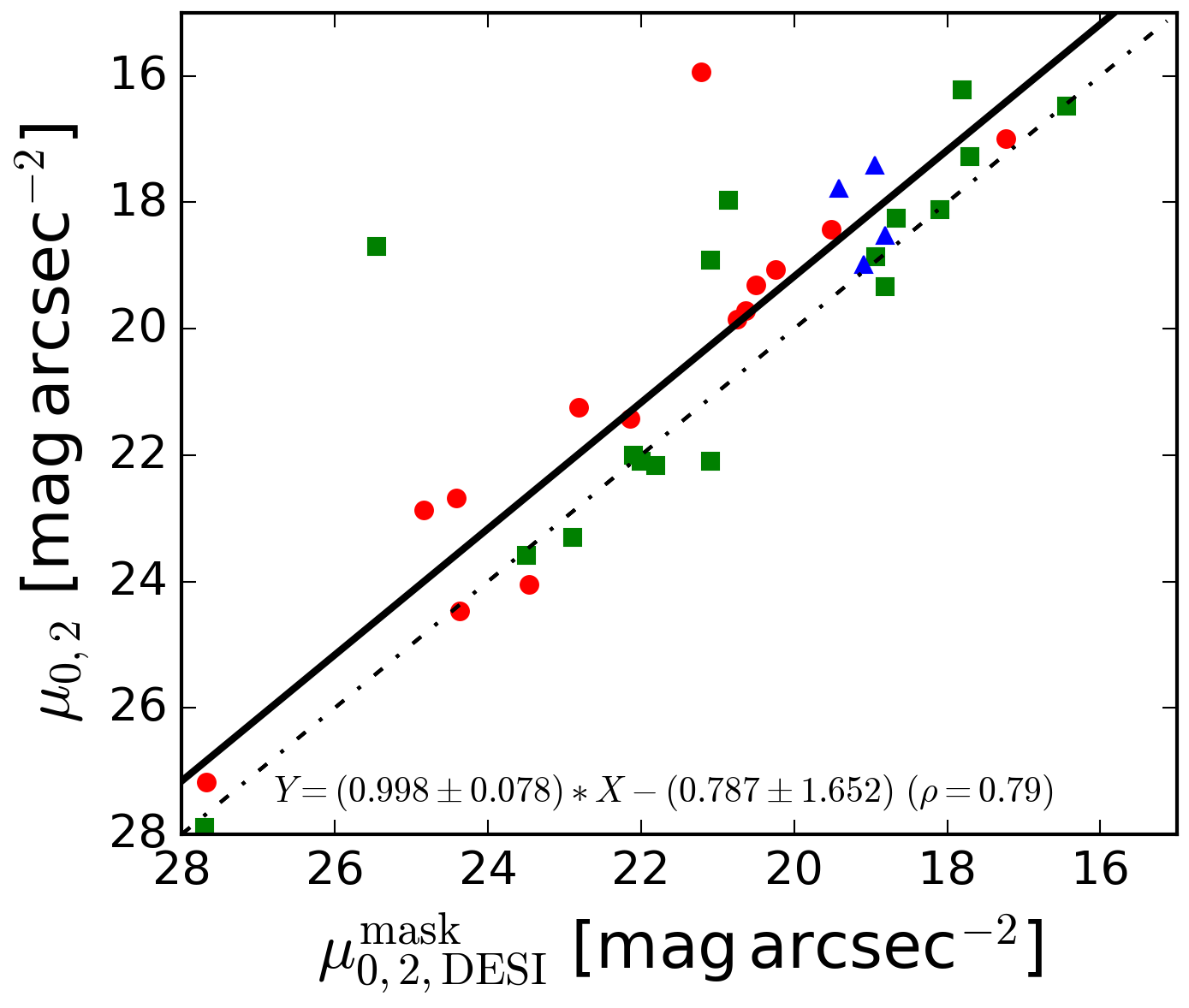}    
    \includegraphics[width = 0.4\linewidth]{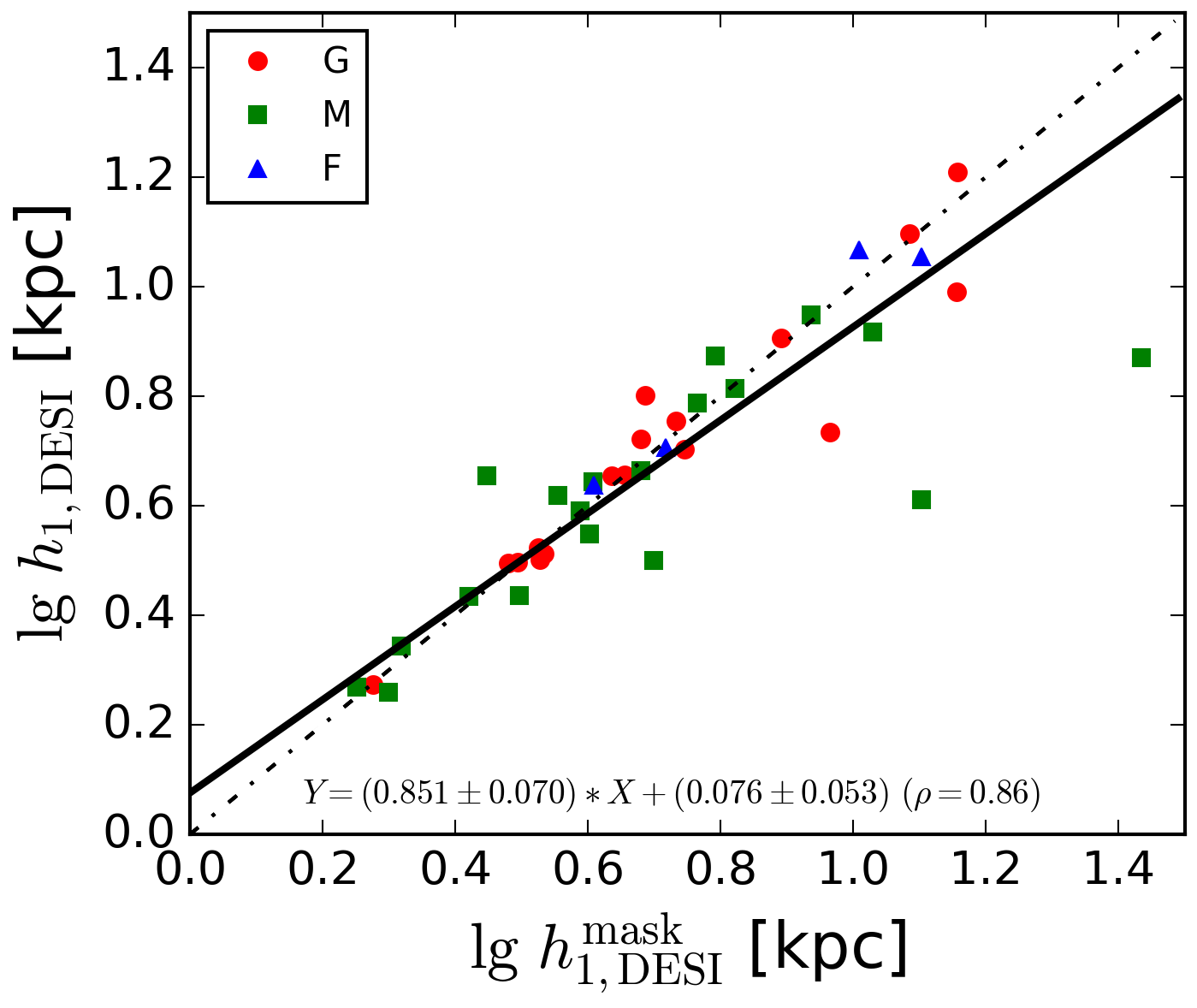}
    \includegraphics[width = 0.4\linewidth]{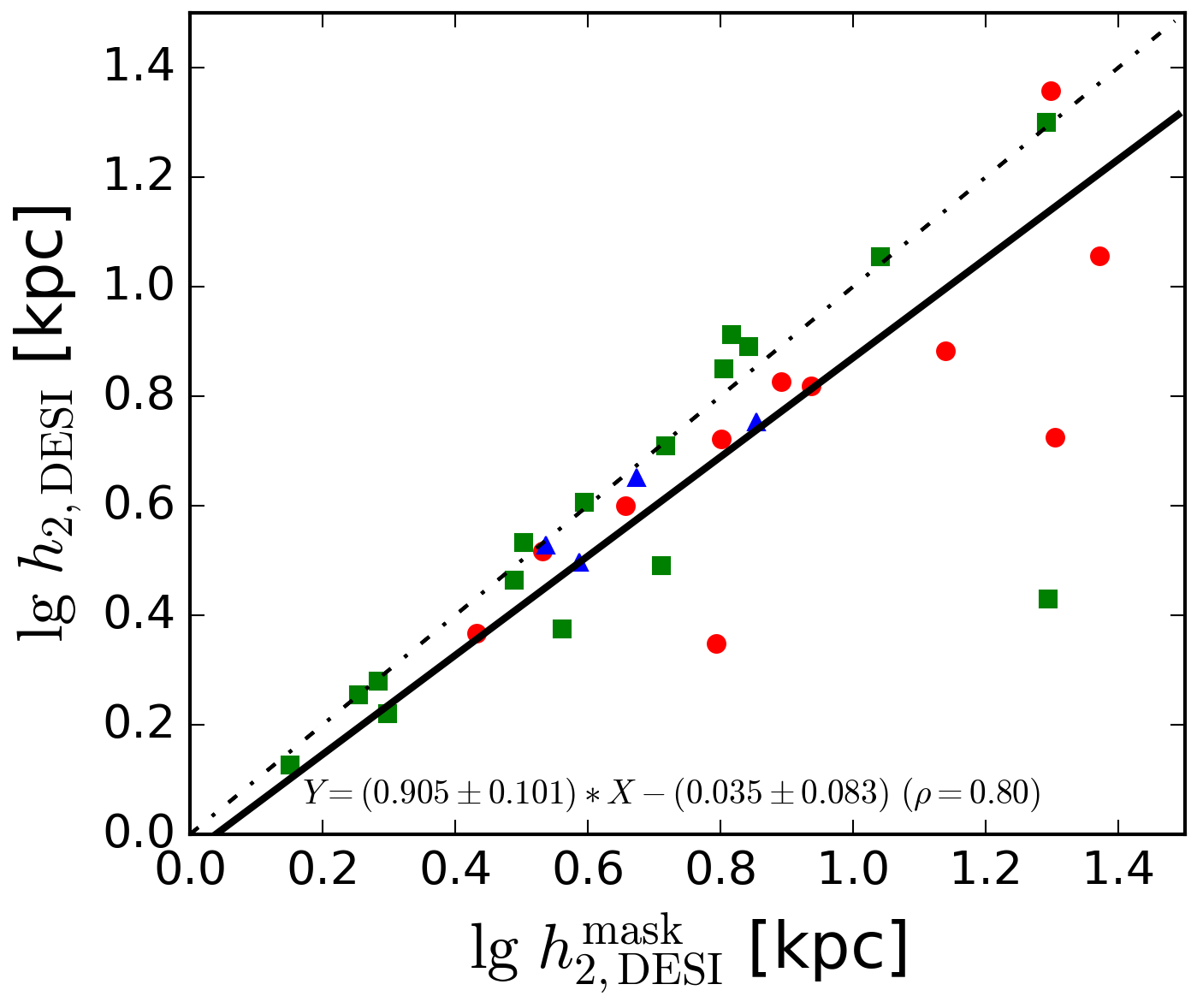}
    \caption{Comparison of the parameters in single exponential and double exponential disc models with (X axis) and without (Y axis) a mask of the spiral arms. The notations are the same as in Fig.~\ref{fig:DES_vs_NUV}.  }
    \label{fig:disks_compar_mask}
\end{figure*}

Since for 67 galaxies in our sample, all main spiral arms have been completely traced, we can assess the accuracy of our fitting for the discs as we mask off the spiral arms. Measuring this effect is of great importance for understanding how spiral structure can influence the disc parameters retrieved in galaxy modelling. Of course, a more accurate decomposition with an additional component for spiral arms is required to explore this question in greater detail (Chugunov et al., in prep.), but this simple approach can give us a general sense of potential biases introduced by spiral arms. Fig.~\ref{fig:disks_compar_mask} illustrates a comparison of the fit parameters for the inner and outer (if present) discs with and without a mask of the spiral arms. As we can see, there may be a systematic overestimation of the first disc central surface brightness for some galaxies when the spiral arms are not masked out (note, however, that most galaxies yet follow the one-to-one relation, which is reflected in the values of the regression line coefficients and their uncertainties). Interestingly, the effect becomes more pronounced for the outer disc in half of the galaxies: $\Delta \mu_{0,1}\simeq0.45$ versus $\Delta \mu_{0,2}\simeq0.85$. The influence of spiral arms on the fitting of the first disc scale length is much less dramatic, but may lead to a systematic underestimation of the second disc scale length for galaxies with bright spiral arms: $\lg\,h_1\simeq0.93\,\lg\,h_1^\mathrm{mask}$ for the inner disc and $\lg\,h_2\simeq0.84\,\lg\,h_2^\mathrm{mask}$ for the outer disc. In other words, the outer discs with unmasked spirals may appear to be less extended than when a mask of the spiral arms is taken into account. The reason for these systematic errors in decomposition without masking of the spiral arms stems from the fact that the spiral arms make some contribution to the total galaxy luminosity (their fraction is typically higher in grand-design galaxies, see figures 19 and 20 in \citetalias{Savchenko_2020}). Therefore, if we neglect the presence spiral arms in our modeling, their luminosity density is ``assigned'' to the stellar disc, which makes the disc brighter and shorter. This effect becomes very significant for estimating the parameters of the outer disc in grand-design galaxies with bright spiral arms. 

As shown in \citet{2017ApJ...845..114G}, to obtain accurate decomposition models for the bulge, nuclear and inner morphological features (lenses, rings, discs, bars) should be taken into account, while outer secondary morphological features (such as rings and spirals) do not influence the retrieved parameters of the bulge significantly. However, according to \citet{2022A&A...659A.141S}, if our goal is to fit a galaxy with a single S\'ersic function (a widely used approach to quantify the morphology of a galaxy, see, e.g. \citealt{2019A&A...622A.132M}), a bright spiral structure can significantly affect the retrieved S\'ersic parameters for galaxies with a prominent bulge. Therefore, accurate photometric decomposition is of great importance for studying the structural properties of galaxies and may be critical for studying the galaxy scaling relations (see e.g. \citealt{2021FrASS...8..157D} and an exhaustive list of references therein), especially if objects with non-axisymmetric features are considered.  

\section{Conclusions}
\label{sec:conclusion}

We have studied the outer spiral structure for a sample of galaxies described in our first paper by \citetalias{Savchenko_2020}. Using GALEX NUV and deep DESI images of the sample galaxies, we have measured the main characteristics of the spiral arms beyond the optical radius. In addition, we explored the structure of the disc profiles using a piecewise linear function to estimate the type of the disc profile and its breaks, along with the disc scale lengths of the regions with different slopes of the surface brightness profile. Our main findings can be summarised as follows.

\begin{enumerate}
    \item The deeper DESI images, analysed in this paper, reveal a slightly more extended spiral structure in grand-design galaxies than was traced in ordinary SDSS frames by \citetalias{Savchenko_2020} (see Fig.~\ref{fig:arms_extent}, right plot). For individual galaxies, the deep imaging revealed a $2-3$ times more extended spiral structure than was traced in \citetalias{Savchenko_2020}. However, multiarmed and flocculent galaxies usually do not demonstrate faint extended spiral arms and cut off at $R_\mathrm{arms}\simeq(1.0\pm0.2)\,R_{25}$. 
    \item In general, there is little difference between the quantitative characteristics of the multiarmed and flocculent galaxies in our sample. However, the spiral arms of grand-design galaxies have, on average, higher maximum azimuthal angles (up to $\Delta \phi\approx12$~rad, see Fig.~\ref{fig:phimaxl}), greater extent (up to $4\,R_{25}$ or $\sim 80$~kpc, see Fig.~\ref{fig:arms_extent}), and larger widths of the outer spiral arms (up to $w\sim0.3\,R_{25}$, see Fig.~\ref{fig:widthall}) than the other two galaxy classes. Each of these differences could be explained by tidal interactions with neighbouring galaxies, which leads to a greater width of the spirals in the periphery of the galaxy. This conclusion is supported by the observation that galaxies in tight spatial environments (group and clusters) have, on average, larger arm widths than those in isolated galaxies (see Fig.~\ref{fig:widthall}, right plot).
    \item The comparison between the extent of spiral arms in the NUV and the optical does not reveal a significant difference between the various classes of spiral structure, as well as between spiral arms with different colors. On average, spiral arms in the NUV and deep optical observations have a similar extent, but less physically extended spirals tend to have smaller radii in the optical than in the NUV (see Fig.~\ref{fig:DES_vs_NUV_arms}).
    \item The pitch angle of galaxies may become systematically greater or lower in the galaxy periphery than within the optical radius, without any dependence on class of spiral structure, type of disc profile, global galaxy morphology, or spatial environment.
    \item According to our analysis of the disc profiles in the sample galaxies, the innermost disc region typically has a shorter disc scale length in the $r$ band than in the NUV: $h_\mathrm{1,DESI}$/$h_\mathrm{1,NUV}=0.74$, in perfect agreement with the literature. Surprisingly, the second disc scale length is, on average, slightly larger in the optical: $h_\mathrm{2,DESI}$/$h_\mathrm{2,NUV}=1.44$. The disc-profile breaks are largely consistent in these wavebands, although surface brightness profiles with completely uncorrelated disc breaks in the NUV and optical are rather common (see Fig.~\ref{fig:DES_vs_NUV} for the bending strength indicator).
    \item Our sample is dominated by Type II discs, whereas pure single-exponential profiles are quite rare. Grand-design galaxies demonstrate a higher fraction of Type III and Type II+III profiles (see Fig.~\ref{fig:profile_types}). According to Fig.~\ref{fig:Rarms_Rbr}, a significant number of galaxies show a tight correlation between the extent of the spiral arms and one of the break radii. Therefore, several mechanisms may be responsible for the existence/absence of disc breaks in galaxies: i) quenching of star formation when entering a galaxy cluster (Type I profile -- no breaks), ii) the influence of the Outer Lindblad Resonance of bars and other types of secular evolution (Type II and Type II+III profiles), iii) interactions of galaxies that can deform the disc and induce galactic feathers or grand-design spiral arms and ``heat'' the galaxy structure (Type III profile).
    \item Ignoring a mask of the spiral structure while fitting the disc profile may affect the derived parameters of the disc: the disc scale length may be slightly underestimated, whereas the central surface brightness may be significantly overestimated: disc models with the masked spiral arms appear to be 0.4--0.8 mag\,arcsec$^{-2}$ fainter than in models where the spirals are not masked out. The effect becomes more pronounced for the outer discs.  
\end{enumerate}

The obtained properties are integral in observing the differences between grand-design galaxies and the other two classes of spirals. Some of the trends detailed above were outlined in \citetalias{Savchenko_2020} using SDSS observations, but became more apparent when deeper imaging allowed the characterisation of spiral features that extend beyond the optical radius. Although the integral characteristics of spiral arms are very similar among the three classes of spiral structure (except for the integral colour and the contribution of the total galaxy luminosity), our study makes clear that at least some fraction of grand-design galaxies should be generated by tidal mechanisms.

\section*{Acknowledgements}
We acknowledge financial support from the Russian Science Foundation (grant no. 22-22-00483).

This research has made use of the NASA/IPAC Extragalactic Database (NED; \url{https://ned.ipac.caltech.edu/}) operated by the Jet Propulsion Laboratory, California Institute of Technology, under contract with the National Aeronautics and Space Administration. This research has made use of the HyperLEDA database (\url{http://leda.univ-lyon1.fr/}; \citealp{Makarov2014}). 

Funding for the Sloan Digital Sky Survey IV has been provided by the Alfred P. Sloan Foundation, the U.S. Department of Energy Office of Science, and the Participating Institutions. SDSS-IV acknowledges
support and resources from the Center for High-Performance Computing at the University of Utah. The SDSS website is www.sdss.org.

SDSS-IV is managed by the Astrophysical Research Consortium for the 
Participating Institutions of the SDSS Collaboration including the 
Brazilian Participation Group, the Carnegie Institution for Science, 
Carnegie Mellon University, the Chilean Participation Group, the French Participation Group, Harvard-Smithsonian Center for Astrophysics, 
Instituto de Astrof\'isica de Canarias, The Johns Hopkins University, Kavli Institute for the Physics and Mathematics of the Universe (IPMU) / 
University of Tokyo, the Korean Participation Group, Lawrence Berkeley National Laboratory, 
Leibniz Institut f\"ur Astrophysik Potsdam (AIP),  
Max-Planck-Institut f\"ur Astronomie (MPIA Heidelberg), 
Max-Planck-Institut f\"ur Astrophysik (MPA Garching), 
Max-Planck-Institut f\"ur Extraterrestrische Physik (MPE), 
National Astronomical Observatories of China, New Mexico State University, 
New York University, University of Notre Dame, 
Observat\'ario Nacional / MCTI, The Ohio State University, 
Pennsylvania State University, Shanghai Astronomical Observatory, 
United Kingdom Participation Group,
Universidad Nacional Aut\'onoma de M\'exico, University of Arizona, 
University of Colorado Boulder, University of Oxford, University of Portsmouth, 
University of Utah, University of Virginia, University of Washington, University of Wisconsin, 
Vanderbilt University, and Yale University.

The Legacy Surveys consist of three individual and complementary projects: the Dark Energy Camera Legacy Survey (DECaLS; NOAO Proposal ID \# 2014B-0404; PIs: David Schlegel and Arjun Dey), the Beijing-Arizona Sky Survey (BASS; NOAO Proposal ID \# 2015A-0801; PIs: Zhou Xu and Xiaohui Fan), and the Mayall z-band Legacy Survey (MzLS; NOAO Proposal ID \# 2016A-0453; PI: Arjun Dey). DECaLS, BASS and MzLS together include data obtained, respectively, at the Blanco telescope, Cerro Tololo Inter-American Observatory, National Optical Astronomy Observatory (NOAO); the Bok telescope, Steward Observatory, University of Arizona; and the Mayall telescope, Kitt Peak National Observatory, NOAO. The Legacy Surveys project is honored to be permitted to conduct astronomical research on Iolkam Du'ag (Kitt Peak), a mountain with particular significance to the Tohono O'odham Nation.

NOAO is operated by the Association of Universities for Research in Astronomy (AURA) under a cooperative agreement with the National Science Foundation.

This project used data obtained with the Dark Energy Camera (DECam), which was constructed by the Dark Energy Survey (DES) collaboration. Funding for the DES Projects has been provided by the U.S. Department of Energy, the U.S. National Science Foundation, the Ministry of Science and Education of Spain, the Science and Technology Facilities Council of the United Kingdom, the Higher Education Funding Council for England, the National Center for Supercomputing Applications at the University of Illinois at Urbana-Champaign, the Kavli Institute of Cosmological Physics at the University of Chicago, Center for Cosmology and Astro-Particle Physics at the Ohio State University, the Mitchell Institute for Fundamental Physics and Astronomy at Texas A\&M University, Financiadora de Estudos e Projetos, Fundacao Carlos Chagas Filho de Amparo, Financiadora de Estudos e Projetos, Fundacao Carlos Chagas Filho de Amparo a Pesquisa do Estado do Rio de Janeiro, Conselho Nacional de Desenvolvimento Cientifico e Tecnologico and the Ministerio da Ciencia, Tecnologia e Inovacao, the Deutsche Forschungsgemeinschaft and the Collaborating Institutions in the Dark Energy Survey. The Collaborating Institutions are Argonne National Laboratory, the University of California at Santa Cruz, the University of Cambridge, Centro de Investigaciones Energeticas, Medioambientales y Tecnologicas-Madrid, the University of Chicago, University College London, the DES-Brazil Consortium, the University of Edinburgh, the Eidgenossische Technische Hochschule (ETH) Zurich, Fermi National Accelerator Laboratory, the University of Illinois at Urbana-Champaign, the Institut de Ciencies de l'Espai (IEEC/CSIC), the Institut de Fisica d'Altes Energies, Lawrence Berkeley National Laboratory, the Ludwig-Maximilians Universitat Munchen and the associated Excellence Cluster Universe, the University of Michigan, the National Optical Astronomy Observatory, the University of Nottingham, the Ohio State University, the University of Pennsylvania, the University of Portsmouth, SLAC National Accelerator Laboratory, Stanford University, the University of Sussex, and Texas A\&M University.

BASS is a key project of the Telescope Access Program (TAP), which has been funded by the National Astronomical Observatories of China, the Chinese Academy of Sciences (the Strategic Priority Research Program "The Emergence of Cosmological Structures" Grant \# XDB09000000), and the Special Fund for Astronomy from the Ministry of Finance. The BASS is also supported by the External Cooperation Program of Chinese Academy of Sciences (Grant \# 114A11KYSB20160057), and Chinese National Natural Science Foundation (Grant \# 11433005).

The Legacy Survey team makes use of data products from the Near-Earth Object Wide-field Infrared Survey Explorer (NEOWISE), which is a project of the Jet Propulsion Laboratory/California Institute of Technology. NEOWISE is funded by the National Aeronautics and Space Administration.

The Legacy Surveys imaging of the DESI footprint is supported by the Director, Office of Science, Office of High Energy Physics of the U.S. Department of Energy under Contract No. DE-AC02-05CH1123, by the National Energy Research Scientific Computing Center, a DOE Office of Science User Facility under the same contract; and by the U.S. National Science Foundation, Division of Astronomical Sciences under Contract No. AST-0950945 to NOAO.

This research has made use of GALEX data obtained from the Mikulski Archive for Space Telescopes (MAST); support for MAST for non-HST data is provided by the NASA Office of Space Science via grant NNX09AF08G and by other grants and contracts (MAST is maintained by STScI, which is operated by the Association of Universities for Research in Astronomy, Inc., under NASA contract NAS5-26555).

%

\section*{Data Availability}

The data underlying this article will be shared on reasonable request to the corresponding author.



\bibliographystyle{mnras}
\bibliography{art} 

\appendix
\section{Characteristics of spiral structure in the sample galaxies}
\label{sec:table_1}
We provide a table with the measured characteristics of spiral structure in the sample galaxies (see details in Sect.~\ref{sec:sp_structure}) with a description of each column. The table itself can be found in the supplementary material in the electronic version of the MNRAS.

\begin{table*}
\caption{Structural parameters of the disc profiles determined for the GALEX NUV and DESI $r$-band data.}
\label{tab:spirals}
\centering
    \begin{tabular}{ll}
    \hline
    \hline    
    Column Name &   Description \\[0.5ex]
    \hline
Object           & Galaxy name\\
$\langle \psi_{r<R_{25}} \rangle$ & Average pitch angle of all spiral arms measured within the optical radius, in degrees \\
$\langle \psi_{r>R_{25}} \rangle$ & Average pitch angle of all measured spiral arms beyond the optical radius, in degrees \\
$R_\mathrm{arms,NUV}$ & Maximum radius of spiral structure in the GALEX NUV image, in kiloparsecs \\
$R_\mathrm{arms,DESI}$ & Maximum radius of spiral structure in the DESI $gr$ combined image, in kiloparsecs \\
$\langle w_{r<R_{25}} \rangle$ & Average width of all spiral arms within the optical radius, in kiloparsecs \\
$\langle w_{r>R_{25}} \rangle$ & Average width of all spiral arms beyond the optical radius, in kiloparsecs  \\
$R_{25}$ & Optical radius from \citetalias{Savchenko_2020}, in kiloparsecs \\
$\Delta \phi$ & Difference between the maximum and minimum azimuthal angles of all spiral arms, in radians\\
Tidal features & Description of tidal structures identified in the vicinity of the target galaxy \\
    \hline\\[-0.5ex]
    \end{tabular}
     \parbox[t]{170mm}{
    \textbf{Notes:}
    The table is available as online material. Values of $\langle \psi_{r<R_{25}} \rangle$, $\langle \psi_{r>R_{25}} \rangle$, $\langle w_{r<R_{25}} \rangle$, $\langle w_{r>R_{25}} \rangle$ and $\Delta \phi$ were measured for the combined masks of spiral arms created for the SDSS (taken from \citetalias{Savchenko_2020}) and DESI (obtained in this study) images. All other characteristic features were obtained for the DESI data.\\
    }  
    
\end{table*}

\section{Characteristics of stellar discs in the sample galaxies}
\label{sec:table_2}
We provide a table with the measured characteristics of stellar discs in the sample galaxies (see details in Sect.~\ref{sec:profiles}) with a description of each column. The table itself can be found in the supplementary material in the electronic version of the MNRAS.

\begin{table*}
\caption{Structural parameters of the disc profiles determined for the GALEX NUV and DESI $r$-band data.}
\label{tab:disks}
\centering
    \begin{tabular}{ll}
    \hline
    \hline    
    Column Name &   Description \\[0.5ex]
    \hline
Object           & Galaxy name\\
Type$_\mathrm{NUV}$     & Disc profile type determined based on GALEX NUV image \\
Type$_\mathrm{DESI}$     & Disc profile type determined based on DESI $r$-band image \\
$BS_\mathrm{NUV}$      & Bending strength of GALEX NUV disc profile \\
$BS_\mathrm{DESI}$      & Bending strength of DESI $r$-band disc profile \\
$h_\mathrm{1,NUV}$     & Radial scale length of the first (inner) disc measured for the GALEX NUV profile, in kiloparsecs  \\
$h_\mathrm{2,NUV}$     & Radial scale length of the second disc measured for the GALEX NUV profile, in kiloparsecs \\
$h_\mathrm{3,NUV}$     & Radial scale length of the third disc measured for the GALEX NUV profile, in kiloparsecs \\
$h_\mathrm{4,NUV}$     & Radial scale length of the fourth disc measured for the GALEX NUV profile, in kiloparsecs \\
$\mu_\mathrm{0,1,NUV}$     & Central surface brightness of the first (inner) disc measured for the GALEX NUV profile, in mag\,arcsec$^{-2}$ \\
$\mu_\mathrm{0,2,NUV}$     & Central surface brightness of the second disc measured for the GALEX NUV profile, in mag\,arcsec$^{-2}$ \\
$\mu_\mathrm{0,3,NUV}$     & Central surface brightness of the third disc measured for the GALEX NUV profile, in mag\,arcsec$^{-2}$ \\
$\mu_\mathrm{0,4,NUV}$     & Central surface brightness of the fourth disc measured for the GALEX NUV profile, in mag\,arcsec$^{-2}$ \\
$R_\mathrm{br,1,NUV}$     & First break radius for the GALEX NUV profile, in kiloparsecs \\
$R_\mathrm{br,2,NUV}$     & Second break radius for the GALEX NUV profile, in kiloparsecs  \\
$R_\mathrm{br,3,NUV}$     & Third break radius for the GALEX NUV profile, in kiloparsecs \\
$h_\mathrm{1,DESI}$     & Radial scale length of the first (inner) disc measured for the DESI $r$-band profile, in kiloparsecs  \\
$h_\mathrm{2,DESI}$     & Radial scale length of the second disc measured for the DESI $r$-band profile, in kiloparsecs \\
$h_\mathrm{3,DESI}$     & Radial scale length of the third disc measured for the DESI $r$-band profile, in kiloparsecs \\
$h_\mathrm{4,DESI}$     & Radial scale length of the fourth disc measured for the DESI $r$-band profile, in kiloparsecs \\
$\mu_\mathrm{0,1,DESI}$     & Central surface brightness of the first (inner) disc measured for the DESI $r$-band profile, in mag\,arcsec$^{-2}$ \\
$\mu_\mathrm{0,2,DESI}$     & Central surface brightness of the second disc measured for the DESI $r$-band profile, in mag\,arcsec$^{-2}$ \\
$\mu_\mathrm{0,3,DESI}$     & Central surface brightness of the third disc measured for the DESI $r$-band profile, in mag\,arcsec$^{-2}$ \\
$\mu_\mathrm{0,4,DESI}$     & Central surface brightness of the fourth disc measured for the DESI $r$-band profile, in mag\,arcsec$^{-2}$ \\
$R_\mathrm{br,1,DESI}$     & First break radius for the DESI $r$-band profile, in kiloparsecs \\
$R_\mathrm{br,2,DESI}$     & Second break radius for the DESI $r$-band profile, in kiloparsecs  \\
$R_\mathrm{br,3,DESI}$     & Third break radius for the DESI $r$-band profile, in kiloparsecs \\
$h_\mathrm{1,DESI}^\mathrm{mask}$     & Radial scale length of the first (inner) disc measured for the DESI $r$-band profile with the spirals masked, in kiloparsecs  \\
$h_\mathrm{2,DESI}^\mathrm{mask}$     & Radial scale length of the second disc measured for the DESI $r$-band profile with the spirals masked, in kiloparsecs \\
$h_\mathrm{3,DESI}^\mathrm{mask}$     & Radial scale length of the third disc measured for the DESI $r$-band profile with the spirals masked, in kiloparsecs \\
$h_\mathrm{4,DESI}^\mathrm{mask}$     & Radial scale length of the fourth disc measured for the DESI $r$-band profile with the spirals masked, in kiloparsecs \\
$\mu_\mathrm{0,1,DESI}^\mathrm{mask}$     & Central surface brightness of the first (inner) disc measured for the DESI $r$-band profile with the spirals masked, in mag\,arcsec$^{-2}$ \\
$\mu_\mathrm{0,2,DESI}^\mathrm{mask}$     & Central surface brightness of the second disc measured for the DESI $r$-band profile with the spirals masked, in mag\,arcsec$^{-2}$ \\
$\mu_\mathrm{0,3,DESI}^\mathrm{mask}$     & Central surface brightness of the third disc measured for the DESI $r$-band profile with the spirals masked, in mag\,arcsec$^{-2}$ \\
$\mu_\mathrm{0,4,DESI}^\mathrm{mask}$     & Central surface brightness of the fourth disc measured for the DESI $r$-band profile with the spirals masked, in mag\,arcsec$^{-2}$  \\
    \hline\\[-0.5ex]
    \end{tabular}
     \parbox[t]{170mm}{
    \textbf{Notes:}
    The table is available as online material. Depending on the adopted galaxy model, some parameters in the table can have NaN values. Central surface brightnesses have not been corrected for Galactic extinction.\\
    }  
    
\end{table*}

\bsp	
\label{lastpage}
\end{document}